\documentclass[11pt]{article}
\usepackage{graphics}
\usepackage{graphicx}

\begin{document}


\makeatletter
\renewcommand{\@biblabel}[1]{\quad#1.}
\makeatother

\hsize=6.15in
\vsize=8.2in
\hoffset=-0.42in
\voffset=-0.3435in

\normalbaselineskip=24pt\normalbaselines

\begin{center}
  {\Large \bf Information thermodynamics: from physics to neuroscience }
\end{center}

\vspace{0.15cm}

\begin{center}
  { Jan Karbowski } 
\end{center}

\vspace{0.05cm}

\begin{center}
  {\it   Institute of Applied Mathematics and Mechanics, \\
    Department of Mathematics, Informatics and Mechanics,  \\
    University of Warsaw, 02-097 Warsaw, Poland}
\end{center}


\vspace{0.1cm}

\begin{abstract}
  This paper provides a perspective on applying the concepts of information
  thermodynamics, developed recently in non-equilibrium statistical physics,
  to problems in theoretical neuroscience. Historically, information and energy
  in neuroscience have been treated separately, in contrast to physics approaches,
  where the relationship of entropy production with heat is a central idea.
  It is argued here that also in neural systems information and energy can be
  considered within the same theoretical framework. Starting from basic ideas
  of thermodynamics and information theory on a classic Brownian particle,
  it is shown how noisy neural networks can infer its probabilistic motion.
  The decoding of the particle motion by neurons is performed with some
  accuracy and it has some energy cost, and both can be determined using
  information thermodynamics. In a similar fashion, we also discuss how neural
  networks in the brain can learn the particle velocity, and maintain that
  information in the weights of plastic synapses from a physical point of view.
  Generally, it is shown how the framework of stochastic and information
  thermodynamics can be used practically to study neural inference, learning,
  and information storing.

\end{abstract}





\vspace{0.1cm}

\noindent Email:  jkarbowski@mimuw.edu.pl


\vspace{2.3cm}






\newpage

\begin{center}
  {\it  ``Earth, air, fire, and water in the end are all made of energy, 
    but the different forms they take are determined by information.
   To do anything requires energy. To specify what is done requires information.''}
\end{center}

\vspace{0.6cm}

\hspace{7cm}  --- Seth Lloyd (2006) \cite{lloyd}

\vspace{1.3cm}

\noindent
{\Large \bf 1. Introduction: Information is physical, so is the brain}

Brain computations require certain amount of energy
\cite{levy,levy2002,laughlin1998,attwell,karbowski2009,karbowski2012}, and the brain
is one of the most metabolically expensive organs in the body \cite{aiello}.
Moreover, the brain energy cost (oxygen and glucose metabolic rates) scales
linearly with the number of neurons \cite{houzel} and sub-linearly with brain
size \cite{karbowski2007}. Every transition in neural circuits, either on a
microscopic or macroscopic scale is associated with some energy dissipation
\cite{nicolis,goldt,karbowski2019,karbowski2021,lynn,deco2023,lefebvre,jk_urban2024}.
Despite all this, a huge majority of neuronal models used in computational
(or theoretical) neuroscience neglects completely the energetic aspect of brain
functioning, as if neural information processing were for free, performed in some
abstract ``mathematical'' hyperspace
(e.g. \cite{dayan,ermentrout,rieke,chaudhuri,marblestone}).
One can argue that brain information processing is relatively cheap
(only about 10-20 Watts for human brain \cite{karbowski2009,aiello}) in comparison
to computations executed by artificial neural networks on semiconductor
hardware (supercomputer involved in Blue Brain Project uses about $4\cdot10^{5}$
Watts for a ``realistic'' simulation \cite{markram,stiefel}).
However, this relative brain energetic efficiency cannot be a justification
for dismissing the metabolic constraints. In fact, handling information in
real neural circuits is energetically demanding, as transmitting 1 bit of
information through a chemical synapse requires about $\sim 10^{5} k_{B}T$ of energy
\cite{laughlin1998}, and acquiring 1 bit by a synapse during synaptic learning
needs a similar amount of $\sim 5\cdot 10^{6} k_{B}T$ \cite{jk_urban2024},
where $k_{B}$ is the Boltzmann constant, and $T$ is the brain temperature.
Importantly, both these energy figures are much larger than the minimum set
by the Landauer limit ($k_{B}T \ln2$; \cite{landauer}).
Most of the energy consumption in the mammalian brain goes for fast electric
signaling, i.e., the generation of action potentials (neural activation) and
synaptic transmission (each of them roughly $2\cdot 10^{8}k_{B}T/min$, for
neuronal firing rates $\sim 5$ Hz)
\cite{attwell,karbowski2009,karbowski2012}, and for fast communication
(spatial traveling of action potentials along axons) \cite{levy2021}.
In contrast, slow chemical signaling associated with synaptic plasticity
(related to learning and memory) requires much less energy, about $4-11\%$ of
the energy cost expanded on the synaptic transmission for low firing rates
\cite{karbowski2019}. These substantial costs are likely the
reason for observing sparse coding in brain networks, where only a small fraction
of neurons and synapses are active at any instant
of time \cite{balasubramanian,niven}. All this suggests that energy is a strong
constraint on neural information processing and storing, and consequently
not all sorts of computations, even theoretically possible, can be implemented
by neural networks in the brain.

The first meaningful connection between physics and neuroscience was made a
long time ago, in 1871, by James Maxwell in his book about heat \cite{maxwell}.
In that book he considered an ``intelligent being'', or ``demon'' that supposedly
breaks the second law of thermodynamics by decreasing the entropy of the physical
system. This thought experiment was a paradox that triggered a confusion
regarding fundamental issues of thermodynamics, and led to a huge literature on
this subject (for reviews see, \cite{leff,maruyama}). The resolution of this
paradox came with the realization that the concept of information also has
to be included in the thermodynamic considerations, i.e., information has
to be treated on equal footing with physical entropy and work \cite{maruyama}.

This realization followed from a seminal observation made by Rolf Landauer
that erasing information always leads to heat dissipation (erasure of 1 bit
causes at least $k_{B}T\ln2$ of energy released into the environment,
\cite{landauer}). In other words, information is physical, since its
storing and processing requires physical hardware and it has to comply
with the laws of physics \cite{bennett,berut,landauer1991,parrondo}.

It seems that one of the main goals of neural networks of any brain is to
accurately estimate the outside signals \cite{rieke,atick,bialek2001,lang,palmer},
which are relevant for the brain, using as little energy as possible
\cite{sterling,still,karbowski2023}.
Based on these estimates, the brain tries to predict the future dynamics of these
signals and to plan action. The outside signals, or inputs coming to brain circuits,
are mostly of stochastic nature, and therefore their estimation and prediction
is additionally complicated and demanding. Given this, it is perhaps not
surprising that the brain has to possess some internal, stable, representation
of the outside world, which can be modified by learning. It is fair
to say, that despite many conceptual developments we have only rudimentary
knowledge (or feeling) of how this representation is created and works.

We can quantify the degree of correlation between outside dynamics and internal
brain dynamics by mutual information, known from Claude Shannon mathematical
theory of communication \cite{shannon}. This concept was brought to neuroscience
by Horace Barlow \cite{barlow} in the late 1950-s. Much later, it was used by
many neuroscientist, starting from Laughlin \cite{laughlin1981},
Atick \cite{atick}, and most notably by Bialek and colleagues
\cite{rieke,bialek,tkacik}.
These approaches aimed at maximization of mutual information, initially
ignoring energetic aspects. Levy and Baxter were likely the first to consider
energetics of information encoding in neural networks \cite{levy,levy2002}.
However, even in these attempts, information and energy were treated as separate
concepts, not directly related to one another.

In contrast, stochastic thermodynamics provides a framework where information
and energy are mutually related, and can be considered and computed within a
single formalism \cite{parrondo,seifert,peliti}. This is because on a micro level,
which includes molecular fluctuations, all relevant degrees of freedom have to
be considered simultaneously. This work provides a perspective on a mutual
connection  between stochastic and information thermodynamics considered
in physics and neural systems, which are intrinsically stochastic due to their
small sizes and strong interactions with fluctuating environment. 
This intrinsic stochasticity is a key ingredient of neurons and synapses
that causes energy dissipation and influences information processing.

The paper is organized as follows. We start, in Section 2, with reviewing
the fundamentals of stochastic dynamics and their relation to stochastic
thermodynamics, with a simple pedagogic example of a Brownian particle moving in
a gravitational field. This example is a basis and leitmotif for next
considerations, which link this stochastic mechanical system with the thermodynamics
of information processing in neural networks. In Section 3, we introduce
the relationship between entropy, information, and energy, in general and
in particular for the Brownian particle from Section 2. Next in Section 4,
we discuss information flow between two coupled subsystems, as a clear example
where entropy production is directly related to information flow, and its
relevance to Maxwell demon. Neural network inferring the velocity of Brownian
particle (or more general stochastic particle) is presented in Section 5,
together with an associated energy cost.
Synaptic plasticity and learning are discussed in Section 6 in the context
of information gain and loss, using stochastic version of BCM model
\cite{bienenstock} together with its energy cost. It is shown here how the
information loss after learning is related to entropy production rate in synapses.
Most of the calculations in Sections 5 and 6 are novel, i.e., standard neural
and synaptic models are analyzed in a new light. 
Finally, in Section 7, we briefly discuss a more general large
scale model of interacting plastic synapses during learning, using Glauber
dynamics \cite{glauber}, in terms of information processing.
We conclude with some general remarks about the relevance of information
thermodynamics to neuroscience.

\vspace{1.5cm}

\noindent
{\Large \bf 2. Stochastic dynamics and thermodynamics.}

\vspace{0.5cm}

\noindent
{\large{\bf {\it Stochastic dynamics.}}}

Small physical systems have internal degrees of freedom that are subject to
fluctuations due to thermal noise (i.e. interactions with the environment
or ``heat bath''). These internal degrees of freedom can be described either
by discrete or continuous time-dependent variables, such as position, velocity,
activity, composition, etc. Let the index $z$ denote an internal variable
(or all relevant internal variables), describing the state of the system,
and let $p(z,t)$ denote probability that the system is in this particular
state at time $t$. Assuming that $z$ follows a Markov process, one can
describe the dynamics of the probability $p(z,t)$ by a master equation
 \cite{vankampen,gardiner}:

\begin{eqnarray}
  \dot{p}(z)= \sum_{z'} \big(w_{zz'}p(z')-w_{z'z}p(z)\big),
\end{eqnarray}         
where $\dot{p}(z)$ denotes the temporal derivative of $p(z)$, and
$w_{zz'}$ is the transition rate for jump from state $z'$ to state $z$.
Here, the variable $z$ can be either discrete or continuous. In the latter
case, one can expand Eq. (1) to obtain the so-called Fokker-Planck equation
(see below).

In the case with a single continuous internal variable $z(t)$, we can
write its stochastic dynamics as the so-called Langevin equation
\cite{vankampen,gardiner}:

\begin{eqnarray}
  \frac{1}{\mu}\dot{z}= F(z,t) + \sigma\eta(t),
\end{eqnarray}  
where $F(z,t)$ is the deterministic generalized force acting on the system,
which can depend on $z$ and on time $t$, and $\mu$ is some parameter which
is inversely proportional to the time scale of the dynamics. The parameter
$\eta(t)$ is the thermal noise acting on the variable $z$, and thus can
be described by a delta-correlated Gaussian random variable, such that
$\langle \eta(t)\rangle= 0$, and $\langle \eta(t)\eta(t') \rangle= \delta(t-t')$.
The parameter $\sigma$ characterizes the magnitude of the thermal noise. 
If $z$ is velocity, then the two parameters, $\mu$ and $\sigma$, are not
independent. In fact, they are mutually coupled by the temperature of the
system $T$ through the relation: $\sigma^{2}= 2k_{B}T/\mu$ \cite{vankampen,gardiner}.
This relation is known as a fluctuation-dissipation theorem, which essentially
means that in the presence of heat bath (characterized by the temperature $T$)
there is some balance between the level of fluctuations in the system
($\sim \sigma$) and the time for which that system approaches equilibrium
($\sim \mu^{-1}$). It should be noted that for neural systems the thermal noise
is not the most important source of noise, at least on the level of the
whole neuron, and thus the temperature does not play a major part
in the considerations of neural activation (see also below).

The dynamics of variable $z$ can be
described equivalently by the dynamics of probability density of the state
variable $z$ in terms of the Fokker-Planck equation as \cite{vankampen,gardiner}

\begin{eqnarray}
\frac{\partial P(z,t)}{\partial t}= - \frac{\partial J(z,t)}{\partial z},
\end{eqnarray} \\
with the probability current (or flux) $J(z,t)$ given by
\begin{eqnarray}
J(z,t)= \mu F(z,t)P(z,t) - \frac{1}{2}(\mu\sigma)^{2}\frac{\partial P(z,t)}{\partial z},
\end{eqnarray}  
where $P(z,t)$ is the probability density for the variable $z$.

In many circumstances, in physical systems one thinks about $z$ as a generalized
position or velocity. In biological systems, $z$ can be either some structural
variable, concentration of some ions or molecules, or system activity.
These are the most common ``state variables'',
although it should be noted that there are no restrictions about what physical
observable a Langevin equation may or may not describe.

For concreteness, we take a specific example of Eq. (2):
a small particle of mass $m$ moving in a gravitational field with some
modulating time-dependent force $F_{0}(t)$ in the fluctuating
environment, with $z(t)$ being the particle velocity $v(t)$. This example will
be our leitmotif in the most part of this paper devoted to neural information
processing and thermodynamics (Sections 5 and 6).
Langevin equation of motion takes a familiar form:

\begin{eqnarray}
  m\dot{v}= -kv + F(t) + \sqrt{2mk}\sigma_{v}\eta
\end{eqnarray}  
where $kv$ is the deterministic part of the resistance force of the environment
with $k$ being the parameter corresponding to the strength of the resistance and
proportional to the size of the particle. The force $F(t)$ is
$F(t)= mg + F_{0}(t)$, with $g$ being the gravitational acceleration,
and $\sigma_{v}$ is the standard deviation (its steady-state value) of the
particle velocity due to the thermal noise $\eta$ acting on it
(random hitting of air particles).
When $F_{0}(t)=0$ the particle is falling freely with velocity dependent
friction and stochastic environmental fluctuations. In this case,
at the steady state ($t \mapsto \infty$) we obtain the
fluctuation-dissipation relation for our moving particle in the form:
$\sigma_{v}^{2}= k_{B}T/m$. This relation indicates that the fluctuations
in the kinetic energy of the particle correspond to one degree of freedom
associated with $k_{B}T/2$ (in 1D). It is instructive to have a sense of
the magnitude of these fluctuation for real particles. For a particle
with the size 0.1 mm and the mass of 1 $\mu$g (assuming the density 1 g/cm$^{3}$)
we obtain $\sigma_{v}= 2 \;\mu$m/s, which is small and cannot be detected by
a naked eye, but it can be observed with a microscope. For a comparison,
for a hundred times greater particle with the size 1 cm and mass 1 g, we get
$\sigma_{v}= 0.002 \;\mu$m/s, which is extremely small.

We can write the Fokker-Planck equation for Eq. (5), and easily solve it,
yielding a Gaussian distribution $P_{v}$ for particle velocity \cite{vankampen}

\begin{eqnarray}
 P_{v}(v,t)=
  \frac{  \exp\Big(- [v- \langle v(t)\rangle]^{2}/2\sigma_{v}^{2}(t) \Big) }
  {\sqrt{2\pi\sigma_{v}^{2}(t)}},
\end{eqnarray} \\  
where $\langle v(t)\rangle$ is the average velocity,
$\langle v(t)\rangle= [v(0) + \int_{0}^{t} dt' e^{\gamma t'}F(t')/m]e^{-\gamma t}$,
with $\gamma= k/m$, and the time dependent variance of velocity is
$\sigma_{v}^{2}(t)= \sigma_{v}^{2}\big(1-e^{-2\gamma t}\big)=
\langle v(t)^{2}\rangle - \langle v(t)\rangle^{2}$.

In the limit when the particle mass is very small, $m \mapsto 0$, we can
neglect the term on the left in Eq. (5), and use the fact that $v= -dx/dt$,
with $x$ being the height of the particle (velocity increases as height decreases).
This corresponds to a standard overdamped approximation \cite{majumdar}, and
then Eq. (5) transforms to

\begin{eqnarray}
\dot{x}= -\frac{F(t)}{k} - \sqrt{2\gamma}\sigma_{x}\eta.
\end{eqnarray}   
This approximation is equivalent to saying that the particle velocity is in a
quasi stationary state, since its dynamic is governed by a fast time constant $\sim m$.
In Eq. (7) we used the rescaling $\sigma_{x}= \sigma_{v}/\gamma$, where $\sigma_{x}$
refers to the standard deviation of particle position $x$. We can also write
the Fokker-Planck equation for the temporal evolution of the distribution
of particle position $P_{x}(x,t)$ and easily solve it, obtaining

\begin{eqnarray}
 P_{x}(x,t)=
  \frac{  \exp\Big(- \frac{[x- \langle x(t)\rangle]^{2}}{4\sigma_{x}^{2}\gamma t} \Big) }
  {\sqrt{4\pi\sigma_{x}^{2}\gamma t}},
\end{eqnarray} \\  
where the average position 
$\langle x(t)\rangle= x(0) - \frac{1}{k}\int_{0}^{t} dt' F(t')$.
Additionally, the variance of particle position is
$\langle x(t)^{2}\rangle - \langle x(t)\rangle^{2}= 2\sigma_{x}^{2}\gamma t$,
which indicates that it is growing proportionally with time, which is a
characteristic of unrestricted Brownian motion. Also, in this limit, equivalent
to the case $\gamma \gg 1$, we have a simple expression for the mean of
particle velocity (as can be easily seen from Eq. (7)),
$\langle v\rangle \approx F(t)/k$. Note that in contrast to
the distribution for particle velocity (Eq. 6), which has a stationary
solution, the distribution for particle position (Eq. 8) never assumes a stationary
form.

\vspace{0.5cm}

\noindent
{\large{\bf {\it Stochastic thermodynamics.}}}

The first law of thermodynamics is essentially the rule for energy conservation.
It turns out that Eq. (2) can be used to derive the first law, as was
realized by Sekimoto \cite{sekimoto1998,sekimoto2010}. The
idea is to treat the state variable in Eq. (2) as generalized velocity,
and introduce additional state variable $u$ representing generalized position,
on which the generalized force also depends, i.e. $F(z,u,t)$, with $z= du/dt$.
Next, we decompose the force $F(z,u,t)$ as:
$F(z,u,t)= -\partial{V(u,t)}/\partial{u} + f_{nc}(z)$, where $V(u,t)$ is
the generalized potential (dependent on $u$ and $t$), and $f_{nc}(z)$ is the
generalized nonconservative force (dependent on velocity $z$). After the
multiplication of both sides of Eq. (2) by $z$ and rearrangement we get
the conservation of generalized ``mechanical energy'' in the form:

\begin{eqnarray}
  \frac{d}{dt} \Big(\frac{1}{2}\mu^{-1} z^{2}
  + V(u,t)\Big)= \frac{\partial V(u,t)}{\partial t}
  + zf_{nc}(z) + \sigma z\eta(t),
\end{eqnarray} \\
where we used the differentiation rule
$dV(u,t)/dt= \frac{\partial V(u,t)}{\partial t}
+  \frac{\partial V(u,t)}{\partial u}\dot{u}$.   
Note that the left hand side of Eq. (9) is the temporal rate of mechanical energy,
represented by $\frac{1}{2}\mu^{-1}z^{2} + V(u,t)$, which is the sum of ``kinetic energy''
(with $\mu^{-1}$ representing the generalized mass) and generalized potential $V(u,t)$.
Equation (9) implies that mechanical energy is lost (or gained) in three different
ways: by temporal changes in the external potential $V$, by the action of nonconservative
force $f_{nc}$, and by the noise ($\sim \eta$). 
The last two factors constitute the heat dissipated to the environment.

In the case of our Brownian particle in the gravitational field we find the law of
energy conservation as

\begin{eqnarray}
  \frac{d E_{mech}}{dt} 
  = -kv^{2} + vF_{0}(t) + \sqrt{2km}\sigma_{v}v\eta,
\end{eqnarray} \\
where $E_{mech}$ is the mechanical energy of the particle, 
$E_{mech}= \frac{1}{2}mv^{2} + mgx$. Averaging this equation over the
distribution of velocities, Eq. (6), yields the mean balance of
energy loss and gain:

\begin{eqnarray}
  \frac{d \langle E_{mech}\rangle}{dt} 
  = -k\langle v^{2}\rangle + \langle v\rangle F_{0}(t) + k\sigma_{v}^{2},
\end{eqnarray} \\
where we used the Novikov theorem \cite{novikov} for determining the average
$\langle v\eta\rangle= \sigma_{v}\sqrt{k/(2m)}$. According to expectations the
mean mechanical energy is lost due to friction (the term -$k\langle v^{2}\rangle$),
and $\langle E_{mech}\rangle$ can be either decreased or increased by the driving
force depending on its sign. But interestingly, $\langle E_{mech}\rangle$ 
is always increased by the presence of thermal fluctuations
(the term $k\sigma_{v}^{2}$). 

Equation (11) in the limit $m\mapsto 0$, equivalent to $\gamma \gg 1$,
and corresponding to the unrestricted Brownian motion [Eqs. (7) and (8)],
takes a simple form

\begin{eqnarray}
  \frac{d \langle E_{mech}\rangle}{dt} \approx
  -k\langle v\rangle^{2} + F_{0}(t)\langle v\rangle
  \nonumber \\  
 \approx - \frac{mg}{k}[mg + F_{0}(t)].
\end{eqnarray} \\
Thus, the rate of mean mechanical energy is negative unless the driving force is negative
(breaking from outside) and sufficiently strong. This means that opposing the
gravitational force can save the mean mechanical energy, or even increase it.
We will come back also to the mechanical energy later in the context of entropy
production and flux.

\vspace{3.5cm}

\noindent
{\Large \bf 3. Entropy, information, and the second law of thermodynamics.}

\vspace{0.5cm}

\noindent
{\large{\bf {\it Entropy, Kullback-Leibler divergence, and information.}}}

For the system with probability $p(z,t)$ described by Eq. (1) one can define
Shannon entropy $S_{z}(t)$ as \cite{shannon,cover}

\begin{eqnarray}
S_{z}(t)= - \sum_{z} p(z,t)\ln p(z,t),
\end{eqnarray} \\  
which is the measure of an average uncertainty about the state of the system,
or the value of the stochastic variable $z$. The larger the entropy, the less is
known about the actual state of the system. The concept of entropy is central
in thermodynamics \cite{leff,maruyama,seifert}, in information theory \cite{cover},
and in the science of complexity \cite{gellmann}.
 
It is worth noting that Shannon entropy is not the only way to define entropy.
There are other definitions of entropy, such as Renyi entropy \cite{renyi,csiszar}
and Tsallis entropy \cite{tsallis}, which are also used in statistical physics
and information theory \cite{amari,liese,gorban}. Shannon entropy in
Eq. (13) is a special case of these more general entropies.

For two different probability distributions describing the same physical
system, i.e., $p(z)$ and $q(z)$, one can define a statistical distance
between them (in fact, it is a pseudo-distance in probability space)
called Kullback-Leibler (KL) divergence
\cite{cover,kullback}

\begin{eqnarray}
  D_{KL}(p||q)=  \sum_{z} p(z)\ln \frac{p(z)}{q(z)}.
\end{eqnarray} \\  
KL divergence is also called the relative entropy, and it is always
non-negative and quantifies the difference between
the distributions $p(z)$ and $q(z)$. Therefore $D_{KL}(p||q)$ can be also
thought as an information gain by observing $p(z)$ distribution in relation
to the baseline distribution $q(z)$. The larger KL divergence, the more
distinct are the two distributions. $D_{KL}$ has many applications in statistical
physics and information theory \cite{cover,kawai}.
We will use it in the following sections for synaptic information gain and loss.

As for the entropy, one can define also other statistical divergences,
such as Renyi and Tsallis divergences \cite{renyi,tsallis}. 
$D_{KL}$ is a special case of these more general divergences.
There exist numerous inequalities relating various types of statistical
divergences \cite{csiszar,sason}, and inequalities relating the rates
of these divergences to stochastic thermodynamics \cite{karbowski_div}.

In the case of two coupled systems described by variables $x$ and $y$, one can
write $z= (x,y)$ and define the joint probability $p_{xy}$, as well as
marginal probability distributions $p_{x}$ and $p_{y}$ for each subsystem
separately. This allows us to introduce the measure of mutual dependency between
the two subsystems, $\ln\frac{p_{xy}}{p_{x}p_{y}}$, which is zero if $x$ and $y$
are independent and nonzero otherwise. The average of this quantity over all
realizations of $x,y$ is called the mutual information $I_{xy}$ between $x$ and
$y$ \cite{cover}

\begin{eqnarray}
  I_{xy}=  \sum_{x,y} p_{xy}\ln \frac{p_{xy}}{p_{x}p_{y}}
  \nonumber \\
\equiv D_{KL}(p_{xy}||p_{x}p_{y}).
\end{eqnarray} \\
Thus the mutual information is the KL divergence between the joint
probability $p_{xy}$ and the product of marginal probabilities
$p_{x}, p_{y}$. Definition in Eq. (15) ensures that mutual information
is always non-negative, and the stronger the dependence between $x$ and $y$,
the larger $I_{xy}$. This is in contrast to entropy, which can be negative
for continuous probability distributions (when summation is replaced by
integration).

From Eq. (15) it follows that the mutual information can be
also represented in terms of entropies \cite{cover}:

\begin{eqnarray}
 I_{xy} = S_{x} - S_{x|y}= S_{y} - S_{y|x},
\end{eqnarray} \\  
where $S_{x|y}$ is the conditional entropy defined as
$S_{x|y}= - \sum_{x,y} p_{xy}\ln p(x|y)$, with $p(x|y)$ denoting the
conditional probability,  $p(x|y)= p_{xy}/p_{y}$, and similarly for
for reverse conditional entropy $S_{y|x}$ and conditional probability
$p(y|x)$.

In recent years information theory in general, and mutual information in
particular, were applied to stochastic processes in different settings.
For example, information theory was used to derive thermodynamic uncertainty
relations \cite{hasegawa}. Mutual information can be helpful in mapping
input trajectory to output trajectory, which is relevant for biochemical
networks \cite{tostevin}. Additionally, mutual information can be used
to discriminate between internal information in the system and the
information coming from external sources \cite{nicoletti2021}, which
may have some relevance in neuroscience. In the latter context, mutual
information was shown to be maximized for critical brain states with
power law distributions of neural activity \cite{fagerholm,shriki}. 
In a broader biological context, it has been argued that evolution
acts to optimize the gathering and representation of information across
many spatial scales \cite{tkacik}.

\vspace{0.5cm}

\noindent
{\large{\bf {\it Entropy production and flow, and the second law.}}}

The temporal derivative of the entropy from Eq. (13) can be decomposed into two
contributions  \cite{schnakenberg,maes2003,esposito}

\begin{eqnarray}
 \frac{dS}{dt}= \dot{S}_{pr} - \dot{S}_{fl},
\end{eqnarray} \\  
where $\dot{S}_{pr}$ is the entropy production rate given by

\begin{eqnarray}
  \dot{S}_{pr}= \frac{1}{2} \sum_{z,z'}(w_{zz'}p_{z'}-w_{z'z}p_{z})
  \ln \frac{w_{zz'}p_{z'}}{w_{z'z}p_{z}},
\end{eqnarray} \\  
and $\dot{S}_{fl}$ is the entropy flow rate given by

\begin{eqnarray}
  \dot{S}_{fl}= \frac{1}{2} \sum_{z,z'}(w_{zz'}p_{z'}-w_{z'z}p_{z})
  \ln \frac{w_{zz'}}{w_{z'z}}.
\end{eqnarray} \\  
The thermodynamic interpretation of $\dot{S}_{fl}$ is that it is
proportional to the heat $\Delta Q$ exchanged with the surrounding
medium, i.e., $\Delta Q= k_{B}T\dot{S}_{fl}\Delta t$, in the short
time interval $\Delta t$. Moreover, the entropy flow can be of
either sign, which reflects the fact that the system can either
gain energy from the environment ($\dot{S}_{fl} < 0$) or dissipate
energy to the environment ($\dot{S}_{fl} > 0$).

The entropy production rate, on the other hand, is always non-negative,
which follows from the fact that the two factors on the right in Eq. (18)
have the same signs, either both positive or negative. Alternatively,
the non-negativity of $\dot{S}_{pr}$ and its lower bound can be
determined from a well known inequality, $\ln(1+x) \ge \frac{x}{1+x}$,
valid for all $x > -1$. Applying this to Eq. (18) leads to

\begin{eqnarray}
  \dot{S}_{pr} \ge \frac{1}{2} \sum_{z,z'}
  \frac{(w_{zz'}p_{z'}-w_{z'z}p_{z})^{2}}{w_{zz'}p_{z'}} \ge 0.
\end{eqnarray} \\  
The fact that $\dot{S}_{pr} \ge 0$ has a tremendous consequence on the
behavior of stochastic objects in the form of the second law of thermodynamics.
In a nutshell, the second law says that the entropy of the isolated physical
system (for which $\dot{S}_{fl}= 0$) never decreases, i.e., 
$dS/dt = \dot{S}_{pr} \ge 0$, which means that disorder of the isolated
system tends to increase in time.

Equations (18) and (19) apply to the general case described by the master equation
(1), however, it is also possible to define $\dot{S}_{pr}$ and $\dot{S}_{fl}$
for continuous stochastic variables described by the Fokker-Planck equation
(3,4). In the latter case we have \cite{tome}

\begin{eqnarray}
  \dot{S}_{pr}= \frac{2}{(\mu\sigma)^{2}} \int dz \; \frac{J(z,t)^{2}}{P(z,t)} \ge 0,
\end{eqnarray} \\  
and
\begin{eqnarray}
  \dot{S}_{fl}= \frac{2}{\mu\sigma^{2}} \int dz \; J(z,t) F(z,t).
\end{eqnarray} 

For the system at steady state, i.e. for $\dot{p}(z,t)= 0$, its entropy
is constant with $dS/dt= 0$, which implies $\dot{S}_{pr}= \dot{S}_{fl}$.
This equality can happen in the two cases. In the first, the
probability flux $J(z,t)= 0$ for continuous variables, and
$w_{zz'}p_{z'}-w_{z'z}p_{z}= 0$ for discrete variables. This situation
describes the so-called detailed balance (where all local probability
fluxes balance each other), which corresponds to the thermodynamic
equilibrium with the environment. In the second case, one can have
nonzero probability flux,  $J(z,t)\ne 0$, and broken detailed balance
$w_{zz'}p_{z'}-w_{z'z}p_{z}\ne 0$. This situation takes place in the so-called
driven systems by outside factors that provide energy and materials for
maintaining the steady state out of equilibrium with the environment.
Such a steady state is called non-equilibrium steady state (NESS)
\cite{seifert,peliti}. All biological systems are out of equilibrium
\cite{nicolis,bennett,mehta}, and many biological processes operate in
a non-equilibrium steady state \cite{lang,seifert}, including neural systems
\cite{karbowski2019,karbowski2021}.

Since at steady state $\dot{S}_{pr}= \dot{S}_{fl}$, one can say roughly that
for any conditions entropy production rate is proportional to the amount
of dissipated energy to the environment. Thus, it is useful to think about
$\dot{S}_{pr}$ as a measure of energy cost of performing non-trivial function
that requires non-equilibrium conditions.

\vspace{0.5cm}

\noindent
{\large{\bf {\it Entropy production and flow for the Brownian particle.}}}

Our Brownian particle falling in the gravitational field represented by Eqs. (5-8)
has entropy (Eq. 13) corresponding to its position distribution $P_{x}(x,t)$ given by
\cite{cover}

\begin{eqnarray}
S_{x}(t)= \frac{1}{2}\ln(4\pi e \sigma_{x}^{2}\gamma t), 
\end{eqnarray} \\  
which grows logarithmically with time. This means that the uncertainty about
the particle position increases weakly with time. However, the entropy rate,
$dS_{x}/dt$ decreases with time as

\begin{eqnarray}
\frac{dS_{x}}{dt}= \frac{1}{2t}.
\end{eqnarray} \\  

The entropy production rate for particle position (the main ``state variable'') can
be found from Eq. (21). For this, we need the probability current $J(x,t)$ (Eq. 4)
for our particle position, which is

\begin{eqnarray}
J(x,t)= \Big(-\frac{F(t)}{k} + \frac{[x-\langle x\rangle]}{2t}\Big) P_{x}(x,t).
\end{eqnarray} \\  
This allows us to find the entropy production rate $\dot{S}_{pr,x}$ in the form

\begin{eqnarray}
\dot{S}_{pr,x}=  \frac{1}{2t} +  \frac{[mg+F_{0}(t)]^{2}}{\gamma k^{2}\sigma_{x}^{2}}.
\end{eqnarray} \\  
Note that when there is no driving force ($F_{0}= 0$), the entropy production rate
decreases all the time to its asymptotic value
$(mg)^{2}/(\gamma k^{2}\sigma_{x}^{2})= m^{3}g^{2}/(k^{3}\sigma_{x}^{2})$.

The entropy flux rate can be quickly found from Eqs. (24) and (26), using the
definition (17). The result is

\begin{eqnarray}
  \dot{S}_{fl,x}=  \frac{[mg+F_{0}(t)]^{2}}{\gamma k^{2}\sigma_{x}^{2}}
  \nonumber \\  
 \approx \frac{k\langle v\rangle^{2}}{k_{B}T}, 
\end{eqnarray} \\
where the second approximate equality comes from using the fluctuation-dissipation
theorem and the approximate equality for the average particle velocity
$\langle v\rangle\approx [mg+F_{0}(t)]/k$ (see Eq. 7).
Thus, in this case the entropy flux is always positive, suggesting energy dissipation
to the environment.

The relationship between the mechanical energy loss and the entropy flux is
(from Eqs. (12) and (27))

\begin{eqnarray}
\frac{d\langle E_{mech}\rangle}{dt} \approx 
-  k_{B}T\dot{S}_{fl,x}
 + F_{0}(t)\langle v\rangle.
\end{eqnarray} \\  
This equation is the manifestation of the first law of thermodynamics, or
equivalently the law of energy conservation. It implies that our (mechanical)
system changes its energy $E_{mech}$ by dissipating heat to the environment
($k_{B}T\dot{S}_{fl,x}$) and by mechanical work performed on the particle by the
external force $F_{0}$. Equation (28) also suggests that the rate of mean
mechanical energy of the Brownian particle is related to the entropy flux rate
for its position, but they are not the same. The energy lost
$d\langle E_{mech}\rangle/dt$ and $\dot{S}_{fl,x}$ are directly proportional only
if $F_{0}= 0$. To conclude, the entropy flux rate is a measure of dissipated energy
(heat) but it does not account for all lost or gained energy of the system.

\vspace{3.5cm}

\noindent
{\Large \bf 4. Information flow between two subsystems and the Maxwell demon.}

\vspace{0.5cm}

In this section we follow closely the main ideas presented in Ref. \cite{horowitz}.
Consider two coupled subsystems $X$ and $Y$ with dynamics of the joint
probability $p_{xy}$ described by the following master equation

\begin{eqnarray}
  \dot{p}_{xy}= \sum_{x'} \big( w^{y}_{xx'}p_{x'y} - w^{y}_{x'x}p_{xy}\big)
  \nonumber \\
 +  \sum_{y'} \big( w^{x}_{yy'}p_{xy'} - w^{x}_{y'y}p_{xy}\big),
\end{eqnarray}         
where $w^{y}_{xx'}$ is the transition rate in the subsystem $X$ from state $x'$
to state $x$, which depends on the actual state $y$ of the second subsystem $Y$
(and similarly for $w^{x}_{yy'}$). The form of the master equation in Eq. (29)
has a bipartite structure, in which simultaneous jumps in the two subsystems
are neglected as much less likely than single jumps.

For this system we can define the rate of mutual information $dI_{xy}/dt$
as \cite{horowitz,allahverdyan}

\begin{eqnarray}
 \frac{dI_{xy}}{dt}= \dot{I}_{x} + \dot{I}_{y},
\end{eqnarray} \\  
where $\dot{I}_{x}= \big[I_{x_{t+dt},y_{t}}-I_{x_{t},y_{t}}\big]/dt$, and
$\dot{I}_{y}= \big[I_{x_{t},y_{t+dt}}-I_{x_{t},y_{t}}\big]/dt$, with $dt\mapsto 0$.
The explicit expressions for $\dot{I}_{x}$ and $\dot{I}_{y}$ are given by
\cite{horowitz}:

\begin{eqnarray}
  \dot{I}_{x} =
 \sum_{x > x', y} \Big(w^{y}_{xx'}p_{x'y} - w^{y}_{x'x}p_{xy}\Big) \ln\frac{p(y|x)}{p(y|x')},
\end{eqnarray}
and
\begin{eqnarray}
  \dot{I}_{y} =
 \sum_{y > y', x} \Big(w^{x}_{yy'}p_{xy'} - w^{x}_{y'y}p_{xy}\Big) \ln\frac{p(x|y)}{p(x|y')}.
\end{eqnarray}\\
The essence of the decomposition in Eq. (30) is that it splits the total rate
of mutual information into two flows of information. The first flow, $\dot{I}_{x}$,
relates to change in mutual information between the two subsystems that is only
due to the dynamics of $X$. The second flow, $\dot{I}_{y}$, is analogous and
relates to $Y$. When $\dot{I}_{x} > 0$, then information is created in the
subsystem $X$ as it monitors the $Y$ subsystem.

In the same manner we can split the rate of entropy of the joint system (X,Y),
i.e. $dS_{xy}/dt$, as well as the joint entropy production rate $\dot{S}_{pr,xy}$
and the joint entropy flux rate $\dot{S}_{fl,xy}$. We have

\begin{eqnarray}
 \frac{dS_{xy}}{dt}= \dot{S}_{x} + \dot{S}_{y},
\end{eqnarray} \\  
where $S_{xy}= - \sum_{x,y} p_{xy}\ln p_{xy}$, and the rates of entropy in each
subsystem $\dot{S}_{x}$ and $\dot{S}_{y}$ are given by

\begin{eqnarray}
  \dot{S}_{x} =
 - \sum_{y} \sum_{x > x'} \Big(w^{y}_{xx'}p_{x'y} - w^{y}_{x'x}p_{xy}\Big) \ln p_{xy},
\end{eqnarray}
and
\begin{eqnarray}
  \dot{S}_{y} =
 - \sum_{x} \sum_{y > y'} \Big(w^{x}_{yy'}p_{xy'} - w^{x}_{y'y}p_{xy}\Big) \ln p_{xy}.
\end{eqnarray}\\
Note that in the particular case of two independent subsystems, we have
$w^{y}_{xx'} \mapsto w_{xx'}$ ($w^{x}_{yy'} \mapsto w_{yy'}$), and the subsystems
entropy rates $\dot{S}_{x}$ and $\dot{S}_{y}$ reduce to
$\dot{S}_{x}= - \sum_{x} \dot{p}_{x}\ln p_{x}$ and
$\dot{S}_{y}= - \sum_{y} \dot{p}_{y}\ln p_{y}$, i.e., with agreement with
the expectations.

Similarly, the joint entropy production $\dot{S}_{pr,xy}$ and entropy flux
$\dot{S}_{fl,xy}$ can be decomposed as

\begin{eqnarray}
\dot{S}_{pr,xy}= \dot{S}_{pr,x} + \dot{S}_{pr,y},
\end{eqnarray} \\  
where

\begin{eqnarray}
  \dot{S}_{pr,x} =
  \sum_{x > x', y} \Big(w^{y}_{xx'}p_{x'y} - w^{y}_{x'x}p_{xy}\Big)
  \ln\frac{w^{y}_{xx'}p_{x'y}}{w^{y}_{x'x}p_{xy}},
  \nonumber \\
  \dot{S}_{pr,y} =
  \sum_{y > y', x} \Big(w^{x}_{yy'}p_{xy'} - w^{x}_{y'y}p_{xy}\Big)
  \ln\frac{w^{x}_{yy'}p_{xy'}}{w^{x}_{y'y}p_{xy}},
\end{eqnarray}\\
and for the entropy flux

\begin{eqnarray}
\dot{S}_{fl,xy}= \dot{S}_{fl,x} + \dot{S}_{fl,y},
\end{eqnarray} \\  
where

\begin{eqnarray}
  \dot{S}_{fl,x} =
  \sum_{x > x', y} \Big(w^{y}_{xx'}p_{x'y} - w^{y}_{x'x}p_{xy}\Big)
  \ln\frac{w^{y}_{xx'}}{w^{y}_{x'x}},
  \nonumber \\
  \dot{S}_{fl,y} =
  \sum_{y > y', x} \Big(w^{x}_{yy'}p_{xy'} - w^{x}_{y'y}p_{xy}\Big)
  \ln\frac{w^{x}_{yy'}}{w^{x}_{y'y}}.
\end{eqnarray}\\
The terms $\dot{S}_{pr,x}, \dot{S}_{pr,y}$ can be interpreted as
local entropy production rates, while $\dot{S}_{fl,x}, \dot{S}_{fl,y}$
are local entropy fluxes. 
As before, $\dot{S}_{pr,x}$ and $\dot{S}_{pr,y}$ are both non-negative,
which means that the second law is valid also in each of the subsystems.

The interesting thing coming from all these equations is that local
entropy productions $\dot{S}_{pr,x}$, $\dot{S}_{pr,y}$ are related
to information flows $\dot{I}_{x}$ and $\dot{I}_{y}$ as
\cite{horowitz}

\begin{eqnarray}
\dot{S}_{pr,x}= \dot{S}_{x} + \dot{S}_{fl,x} - \dot{I}_{x},
  \nonumber \\
\dot{S}_{pr,y}= \dot{S}_{y} + \dot{S}_{fl,y} - \dot{I}_{y}.  
\end{eqnarray} \\  
These equations imply that local entropy balance involves both energy
dissipation ($\dot{S}_{fl,x}, \dot{S}_{fl,y}$) and the flow of information
($\dot{I}_{x}, \dot{I}_{y}$). Consequently, energy and information are
mutually coupled, and one influences the other. Equalities (40) provide
an important link between information processing and its energy cost.

How do the results represented by Eq. (40) relate to the Maxwell demon?
Although the quantity $\dot{S}_{pr,x}$ always satisfies $\dot{S}_{pr,x} \ge 0$,
the sum $\dot{S}_{x} + \dot{S}_{fl,x}$ can be negative if the information
flow $\dot{I}_{x} < 0$. Thus, from a local point of view of the subsystem
$X$, its visible ``entropy production'' (i.e., $\dot{S}_{x} + \dot{S}_{fl,x}$)
can be negative if the presence of the $Y$ subsystem is neglected. This
seems like a violation of the second law (requiring positive entropy
production rate), and it is closely related to the Maxwell demon thought
experiment. Obviously, the inclusion of the information flow term 
$\dot{I}_{x}$ in the local entropy production solves the paradox.

\vspace{1.5cm}

\noindent
{\Large \bf 5. Neural inference.}

In this section we consider a simple model of how neurons estimate an external
signal. We will discuss this model in terms of information processing as well
as thermodynamics.

Neurons in visual cortex selectively respond to different velocities
of a moving stimulus \cite{rieke,rodman}. Generally, each neuron has a preferred
velocity to which it responds in the form of elevated firing rate
(it is called a tuning curve, see e.g. \cite{dayan,rieke}). 
Thus a single neuron is unable to estimate (decode) the velocity of
the moving stimulus, because it reacts only to a very small range
of velocities. However, a large population of neurons can do it,
although with some accuracy. Below, we consider how such a decoding
can take place. In the example below, which is mostly a ``thought
experiment'', the moving stimulus should be a particle with a substantial
size and velocity to be detectable by visual neurons.
Typical Brownian particles are too small and too slow to be directly
observable by the mammalian visual system. To make them observable,
a magnifying instrument such as a microscope is needed. Thus, one
can think about the moving stimulus below as a magnified Brownian
particle from Section 2, or alternatively, as a macroscopic object
moving stochastically, e.g., due to strong stochastic force $F_{0}$ not
related to thermal fluctuations of the environment. The analysis below
is independent of either choice.

The model we use is a stochastic version of the deterministic model called
linear recurrent network for interacting neurons
(see Eq. (7.17) in \cite{dayan}). In this model, activity or firing rate $r_{i}$
(number of action potentials or spikes per time unit) of a single neuron labeled
as $i$ in the visual cortex can be represented as

\begin{eqnarray}
  \dot{r}_{i}= -\frac{(r_{i}-c_{i}(v))}{\tau_{n0}}
   + \frac{1}{N}\sum_{j} w_{ij}r_{j}
  + \sqrt{\frac{2\sigma_{r0}^{2}}{\tau_{n0}}}\eta_{i}(t),
\end{eqnarray} \\  
where $w_{ij}$ is the synaptic weight (or strength) characterizing the
magnitude of synaptic transmission coming from neuron $j$, and $i=1,2,...,N$,
with $N$ number of neurons in the network (here $w_{ij}$ are in units of inverse of time).
Since the majority of synapses in the cortex of mammals is excitatory
(about $80-90\%$; \cite{braitenberg,karbowski2014}), the weights $w_{ij}$ are
assumed positive, which implies that the steady state average values of $r_{i}$ are
all positive. The parameter $\tau_{n0}$ is the time constant of the single neuron
dynamics related to changes in its firing rate, and $\sigma_{r0}$ is the standard
deviations of the Gaussian noise $\eta_{i}$ related to firing rate fluctuations.
The function $c_{i}(v)$ is the sensory input coming to neuron $i$, which is discussed
below. The activity of the neuron $i$ is a compromise between this sensory input
and the synaptic contributions coming from other neurons in the network.
It should be also clearly stated that the noise term in Eq. (41) is not of thermal
origin. Microscopic thermal fluctuations present in synapses and different ion
channels have only marginal influence on neural activity, since their numbers
for a typical cortical neuron are very large (small variance), although there
are some exceptions (see, \cite{faisal}).
More important are fluctuations caused by unreliable sensory signal and
unpredictable synaptic transmission (probabilistic neurotransmitter release),
the latter caused by low numbers of signaling molecules involved
\cite{renart}.

Before we go further, let us talk about the range of validity of Eq. (41).
First, both the linear term associated with synaptic interactions and the
additive noise can occasionally make the firing rates $r_{i}$
to be negative, which is obviously wrong (even if all synaptic weights are positive).
However, this can happen only transiently, especially in the limit of weak noise.
Moreover, the steady-state average values of firing rates are always positive, since
on average the term $\sum_{j} w_{ij}r_{j}$ is positive.
This means that the linear approximation is a ``reasonable'' approximation, and we
use it primarily because such a linear model can be analytically analyzed, revealing
some generic features. 
Second, the time constant $\tau_{n0}$ in Eq. (41) cannot be too small. It must be
significantly larger than a time constant related to synaptic transmission (5 msec
and 120 msec, related to AMPA and NMDA synaptic receptors),
such that synaptic currents assume quasi-stationary values \cite{dayan}.
In what follows, i.e., the analysis of the dynamics and information aspects of
this model is a novel calculation.

The sensory input $c_{i}(v)$ received by neuron $i$ is in this particular settings
also called the tuning curve for neuron $i$. It can be approximated by a Gaussian as
(see Eq. (3.28) in \cite{dayan})

\begin{eqnarray}
c_{i}(v)= r_{m}\exp\Big[-\frac{(v-u_{i})^{2}}{2\epsilon^{2}}\Big],
\end{eqnarray}  
where $r_{m}$ is the maximal firing rate in response to the visual stimulus
(the same for all neurons in the network),
$u_{i}$ is the preferred velocity for the neuron $i$, and $\epsilon$ characterizes
the maximal deviation from the preferred velocity for which neurons are still (weakly)
activated. We take $\epsilon$ to be small, i.e., typically $\epsilon/u_{i} \ll 1$.
Note that for $v= u_{i}$ we have $c_{i}(v)= r_{m}$, while for
$v= u_{i}\pm 2\epsilon$ we have $c_{i}(v)= 0.14r_{m}$.
Eq. (41) indicates that the neuron adjusts dynamically to the changes in the stimulus
(in its sensitivity range represented by $c_{i}(v)$) and in the synaptic input
coming from other neurons.

Since the decoding of stimulus velocity is a collective process, we define
a population average of all neural activities, denoted as $\bar{r}$, and
defined as $\bar{r}= (1/N)\sum_{i=1}^{N} r_{i}$. Consequently, the dynamic
of the population average neural activity $\bar{r}$ can be represented
as 

\begin{eqnarray}
 \dot{\bar{r}}= -\frac{[\bar{r} - \kappa(\bar{w})\bar{c}(v)]}{\tau_{n}}
  + \sqrt{\frac{2\sigma_{r}^{2}}{N\tau_{n}}}{\bar\eta}(t),
\end{eqnarray} \\  
where we made a mean-field type approximation 
$(1/N^{2})\sum_{i,j} w_{ij}r_{j}\approx \bar{w}\bar{r}$, where $\bar{w}$ is the
population average synaptic weight, i.e., 
$\bar{w}= (1/N^{2})\sum_{i}\sum_{j} w_{ij}$.
We assume that $\bar{w} > 0$, which follows from the fact that the majority of
synapses are excitatory \cite{braitenberg,karbowski2014}. 
The term $\bar{c}(v)$ is population average tuning curve,  
$\bar{c}(v)= (1/N)\sum_{i=1}^{N} c_{i}(v)$, given by (see Appendix A)

\begin{eqnarray}
  \bar{c}(v)\approx \frac{r_{m}\epsilon}{\alpha}\exp\Big(-\frac{v^{2}}{2\alpha^{2}}\Big) 
    \nonumber \\
  \approx \frac{r_{m}\epsilon}{\alpha}\Big[1 - \frac{v^{2}}{2\alpha^{2}} + O(\alpha^{-4})\Big],
\end{eqnarray} \\  
where $\alpha$ is the velocity range to which neurons respond.
The approximate equality in Eq. (44) follows from the fact that $\alpha$ is generally
large, i.e. $\alpha \gg 1$, and we will use that approximation in the calculations
below. The parameter $\kappa(\bar{w})$ is the network enhancement
factor given by

\begin{eqnarray}
  \kappa(\bar{w})= \frac{1}{(1-\bar{w}\tau_{n0})},
\end{eqnarray} \\
since for $\bar{w}\mapsto \tau_{n0}^{-1}$ the parameter $\kappa(\bar{w}) \mapsto \infty$
(obviously, we must assume that $\bar{w} < \tau_{n0}^{-1}$).
The parameter $\tau_{n}$ is the effective time constant of the neural
population dynamics $\tau_{n}= \kappa(\bar{w})\tau_{n0}$,
and $\bar{\eta}$ is the population averaged noise, i.e.,
$\bar{\eta}= (1/\sqrt{N})\sum_{i=1}^{N} \eta_{i}$ with zero mean and unit
variance, with $\sigma_{r}$ being the effective standard deviation of the
noise in the network,
i.e.  $\sigma_{r}= \sqrt{\kappa(\bar{w})}\sigma_{r0}$.
Note that the main effect of the network interactions on population
dynamics, as compared to the single neuron dynamics, is to significantly
enhance the tuning curve, the time constant, and the standard deviation.

Eq. (43) corresponds to the time-dependent distribution of mean neural activity
conditioned on stimulus velocity $v$, $\rho(\bar{r}|v,t)$, in the form
\cite{vankampen}

\begin{eqnarray}
  \rho(\bar{r}|v,t)=
  \frac{  \exp\Big(- N[\bar{r}- \langle \bar{r}(v,t)\rangle_{\rho(\bar{r}|v)}]^{2}/
    2\sigma_{r}^{2}(t) \Big) }
  {\sqrt{2\pi\sigma_{r}^{2}(t)/N}},
\end{eqnarray} \\  
where $\sigma_{r}^{2}(t)= \sigma_{r}^{2}(1-e^{-2t/\tau_{n}})$, and
$\langle \bar{r}(v,t)\rangle_{\rho(\bar{r}|v)}$ is the stochastic average of
the population mean of neural activity over the conditional distribution
$\rho(\bar{r}|v,t)$, i.e.
$\langle \bar{r}(v,t)\rangle_{\rho(\bar{r}|v)}
= \int \; d\bar{r} \rho(\bar{r}|v,t) \bar{r}$. 
The latter can be found quickly by averaging of Eq. (43) over noise, and then
by finding its time dependent solution. The result
is

\begin{eqnarray}
  \langle \bar{r}(v,t)\rangle_{\rho(\bar{r}|v)}=
  \bar{r}(0)e^{-t/\tau_{n}} + \frac{\epsilon\kappa r_{m}}{\alpha}(1-e^{-t/\tau_{n}})
    \nonumber \\
    - \frac{\epsilon\kappa r_{m}}{2\tau_{n}\alpha^{3}}e^{-t/\tau_{n}}
    \int_{0}^{t} \; dt' \; e^{t'/\tau_{n}} v^{2}(t')  +  O(\alpha^{-5}),
\end{eqnarray} \\
where $\bar{r}(0)$ is the initial mean neural activity.
This equation indicates that the outside stimulus modulates the collective neural
activity only weakly (on the order of $\sim \alpha^{-3}$). In addition,
neurons respond to the stimulus with some delay governed by the effective time
constant $\tau_{n}$.

\vspace{0.5cm}

\noindent
{\large\bf {\it Mutual information between neural activities and the stimulus.}}

The degree of correlations between the neural collective activity and the stimulus
velocity is quantified by mutual information $I(\bar{r},v)$ as

\begin{eqnarray}
I(\bar{r},v)= 
\langle \ln\rho(\bar{r}|v)\rangle_{P(\bar{r},v)} -
\langle \ln\rho(\bar{r})\rangle_{\rho(\bar{r})},
\end{eqnarray} \\
where $\rho(\bar{r})$ is the distribution of neural activities, and
averaging in the first term is performed over the joint probability density
$P(\bar{r},v)$ of collective neural activity $\bar{r}$ and stimulus velocity $v$,
with $P(\bar{r},v)= \rho(\bar{r}|v)P(v)$. The distribution $\rho(\bar{r})$ is
found by marginalizing the joint distribution $P(\bar{r},v)$ over velocities $v$.
The result of this procedure is (up to order $\alpha^{-3}$)

\begin{eqnarray}
  \rho(\bar{r},t)\approx 
  \frac{  \exp\Big(- \frac{N}{2\sigma_{r}^{2}(t)}
      \big[\bar{r}- r_{0} + \frac{r_{1}}{\alpha^{3}}\int_{0}^{t}
      dt' e^{t'/\tau_{n}} \langle v^{2}(t')\rangle_{P(v)}\big]^{2} \Big) }
  {\sqrt{2\pi\sigma_{r}^{2}(t)/N}},
\end{eqnarray} \\  
where $r_{0}$ and $r_{1}$ are the stimulus independent and the stimulus dependent
collective neural activities

\begin{eqnarray}
r_{0}= \bar{r}(0)e^{-t/\tau_{n}} + \frac{\epsilon\kappa r_{m}}{\alpha}(1-e^{-t/\tau_{n}}),
    \nonumber \\
r_{1}=  \frac{\epsilon\kappa r_{m}}{2\tau_{n}}e^{-t/\tau_{n}}.
\end{eqnarray} \\
Since both distributions $\rho(\bar{r})$ and $\rho(\bar{r}|v)$ are Gaussian,
the mutual information $I(\bar{r},v)$ can be calculated easily as

\begin{eqnarray}
  I(\bar{r},v)\approx \frac{N(\epsilon\kappa r_{m})^{2}}{8\alpha^{6}\sigma_{r}^{2}(t)}
  e^{-2t/\tau_{n}}  \int_{0}^{t} dt_{1}\int_{0}^{t} dt_{2} e^{(t_{1}+t_{2})/\tau_{n}}
    \nonumber \\
  \times \Big[ \langle v^{2}(t_{1}) v^{2}(t_{2})\rangle_{P(v)}
    - \langle v^{2}(t_{1})\rangle_{P(v)}\langle v^{2}(t_{2})\rangle_{P(v)} \Big]
  + O(\alpha^{-8}).
\end{eqnarray} \\  
This equation shows that the mutual information between neural collective activity and
the stimulus velocity is proportional to the averaged temporal auto-correlations of the
velocity square. Moreover, the larger the number of neurons $N$ decoding the stimulus and
the larger the network enhancement factor $\kappa$, the higher the mutual information.
This clearly indicates that the effect of the network is a key ingredient for the accurate
decoding of information from the outside world.

It is also interesting to see the effect of the timescale associated
with variability in the stimulus velocity on the mutual information $I(\bar{r},v)$
in Eq. (51). Assuming that the temporal auto-correlations of the stimulus
velocity square are characterized by time constant $\tau_{c}$, i.e. that
they decay exponentially as
$\langle v^{2}(t_{1}) v^{2}(t_{2})\rangle_{P(v)}
- \langle v^{2}(t_{1})\rangle_{P(v)}\langle v^{2}(t_{2})\rangle_{P(v)}
= C_{0}e^{-|t_{1}-t_{2}|/\tau_{c}}$,
where $C_{0}$ is some constant, we find that (see Appendix B)

\begin{eqnarray}
  I(\bar{r},v)_{t\mapsto\infty}\approx
  \frac{N(\epsilon\kappa r_{m})^{2}}{8\alpha^{6}\sigma_{r}^{2}}
 \frac{C_{0}\tau_{n}^{2}\tau_{c}}{(\tau_{n}+\tau_{c})}
  + O(\alpha^{-8}).
\end{eqnarray} \\  
This implies that for very fast variability in the stimulus velocity
($\tau_{c} \mapsto 0$) the mutual information between the stimulus and
the neural activity is close to 0. Consequently, neurons in this limit
cannot track the particle velocity at all. However, as the stimulus
variability slows down ($\tau_{c}$ grows), the mutual information increases
and saturates at $\tau_{c} \gg \tau_{n}$. This means that, in this limit,
neurons can decode the stimulus optimally. In general, this result
shows that timescale separation is important for the quality of
neural inference, with a preference for slower stimuli, which agrees
with the general results obtained in \cite{nicoletti2024}.

\vspace{0.5cm}

\noindent
{\large\bf {\it Energy cost of decoding the stimulus.}}

Guessing the actual value of the stimulus by neural network is not free of cost.
In fact, it requires some amount of energy that neurons have to use to perform
that function well. The energy used by neurons can be estimated by calculating the
entropy production rate, with the help of Eq. (21), with probability density
represented by conditional distribution for collective neural activity given by
Eq. (46). We find the conditional entropy production rate $\dot{S}_{\rho(\bar{r}|v)}$
(conditioned on the stimulus velocity $v$) of neural activity as

\begin{eqnarray}
\dot{S}_{\rho(\bar{r}|v)}= \frac{1}{\tau_{n}}
\Big( \frac{e^{-2t/\tau_{n}}}{(e^{2t/\tau_{n}}-1)}
    + \frac{N[\langle \bar{r}(t)\rangle_{\rho(\bar{r}|v)} - \kappa\bar{c}(v)]^{2}}
    {\sigma_{r}^{2}} \Big).
\end{eqnarray} \\  
This formula indicates that the higher the discrepancy between the population
averaged tuning curve and the averaged neural activity, the larger the
entropy production rate of neurons. In other words neurons make an energetic effort
in keeping track of the actual particle velocity.

More explicit formula for the entropy production for longer times ($t \gg t_{n}$),
after transients are gone, is

\begin{eqnarray}
  \dot{S}_{\rho(\bar{r}|v)}\approx \frac{N(\epsilon\kappa r_{m})^{2}}
      {4\alpha^{6}\sigma_{r}^{2}\tau_{n}}
      \Big( v^{2}(t) - \frac{e^{-t/\tau_{n}}}{\tau_{n}}\int_{0}^{t} \;dt'
       e^{t'/\tau_{n}} \langle v^{2}(t')\rangle_{P(v)} \Big)^{2},
\end{eqnarray} \\  
which implies that $\dot{S}_{\rho(\bar{r}|v)}$ is proportional to fluctuations
of the square of velocity around its delayed average. Thus, for stationary
stimulus velocity, its tracking by neurons is essentially energetically costless
(neurons however use energy for other biophysical processes,
\cite{attwell,karbowski2009,karbowski2012,karbowski2019}).
Note also that the prefactor in Eq. (54) is the same as that in Eq. (51) for
the mutual information between neural activities and the stimulus velocity.
This means that gaining information about the outside signals requires
proportionally large supply of energy, i.e., better prediction needs
proportionally more energy.

\vspace{1.5cm}

\noindent
{\Large \bf 6. Stochastic dynamics of synaptic plasticity: learning and memory
storage}

Synaptic weights are not fixed but change in the neural network, although
much slower than neural electric activities. Synaptic plasticity is the
mechanism with which synaptic weights change, and it is responsible for
learning and memory formation in neural systems \cite{dayan,kandel,bourne,takeuchi,poo}.
The model analyzed in this Section is a novel extension and modification of the model
analyzed in \cite{karbowski2021}.

\vspace{0.5cm}

\noindent
{\large\bf {\it Dynamics of synaptic weights.}}

One of the most influential and important models of synaptic plasticity
is the so-called BCM model \cite{bienenstock}, which was used for understanding
the development of the mammalian visual cortex. It is an extension
of the Hebb idea that connections between simultaneously activated pre- and
post-synaptic neurons become stronger, but constructed in such a way that
the synaptic weights stabilize at some level, without catastrophic run-away
as it takes place for a classic Hebb's rule \cite{dayan}. The BCM plasticity
rule, which was originally a deterministic rule, was extended to a stochastic
rule by the author in \cite{karbowski2021}, because synaptic plasticity
is stochastic in nature \cite{meyer,statman}. In the case of a given post-synaptic
neuron with activity $r$, which receives $N_{s}$ synaptic inputs from neurons
with activities $f_{i}$ ($i=1,...,N_{s}$), the stochastic BCM rule takes the
following form \cite{karbowski2021}

\begin{eqnarray}
\frac{dw_{i}}{dt} = \lambda f_{i}r(r-\theta) - \frac{w_{i}}{\tau_{w}}
+ \frac{\sqrt{2}\sigma_{w}}{\sqrt{\tau_{w}}} \xi_{i}      \\
\tau_{\theta}\frac{d\theta}{dt} = -\theta + \beta r^{2},
\end{eqnarray}\\
where $w_{i}$ is the synaptic weight (proportional to the number of
receptors on a synaptic membrane) related to electric conductance of
signals coming from pre-synaptic neuron $i$, $\lambda$ is the
amplitude of synaptic plasticity controlling the rate of change of synaptic
weight, $\tau_{w}$ is the synaptic time constant controlling the weight decay
duration, $\theta$ is the homeostatic variable the so-called sliding threshold 
(adaptation for plasticity) related to an interplay of LTP and LTD
(respectively, long-term potentiation and long-term depression; \cite{dayan})
with the time constant $\tau_{\theta}$, and $\beta$ is the coupling intensity of
$\theta$ to the post-synaptic firing rate $r$. The parameter $\sigma_{w}$ is the
standard deviation of weights due to stochastic intrinsic fluctuations in
synapses, which are represented as Gaussian white noise $\xi_{i}$ with
zero mean and Delta function correlations,
i.e., $\langle\xi_{i}(t)\rangle_{\eta}= 0$ 
and $\langle\xi_{i}(t)\xi_{j}(t')\rangle_{\eta}= \delta_{ij}\delta(t-t')$ 
\cite{vankampen}. Eqs. (55) and (56) correspond to plastic synapses located
on a single neuron. We consider this example, because it is easier to analyze
than the whole network of neurons.

It is often assumed that $\tau_{\theta}/\tau_{w} \ll 1$, and then
the homeostatic variable achieves a steady-state on the time scale for changes
in synaptic weights, i.e. $d\theta/dt\approx 0$. This means that for long times
we have approximately $\theta\approx \beta r^{2}$, and consequently the BCM
rule takes a simple (one equation) form

\begin{eqnarray}
\frac{dw_{i}}{dt} = \lambda f_{i}r^{2}(1-\beta r) - \frac{w_{i}}{\tau_{w}}
+ \frac{\sqrt{2}\sigma_{w}}{\sqrt{\tau_{w}}} \xi_{i}.
\end{eqnarray}\\

As we saw in the previous section, the neural network function is determined
primarily by the collective dynamics of neurons and synapses. For that reason,
it makes sense to consider also the dynamics of the population averaged synaptic
weight. In this case it is not the population average of all synapses in the
network, but rather the population average of synapses on a single neuron,
i.e., $\bar{w}= (1/N_{s})\sum_{i} w_{i}$. Summing both sides of Eq. (57) with the
rescaling factor $N_{s}$, we obtain the population averaged dynamics of $\bar{w}$

\begin{eqnarray}
  \frac{d\bar{w}}{dt} = \lambda \bar{f}r^{2}
 (1-\beta r) - \frac{\bar{w}}{\tau_{w}}
+ \frac{\sqrt{2}\sigma_{w}}{\sqrt{N_{s}\tau_{w}}} \bar{\xi},
\end{eqnarray}\\
where $\bar{f}= (1/N_{s})\sum_{i} f_{i}$, and
$\bar{\xi}= (1/\sqrt{N_{s}})\sum_{i} \xi_{i}$.
Moreover, the neural activity is much faster than the synaptic dynamics (seconds
vs. minutes), i.e. $\tau_{n0}/\tau_{w} \ll 1$. Hence the neural dynamics also reaches
quasi stationary state on the time scales $\sim \tau_{w}$, and it can be approximated
by (from Eq. (41)) 

\begin{eqnarray}
r \approx c(v) + \tau_{n0}\bar{f}\bar{w},
\end{eqnarray} \\
where we used a mean-field expression $(1/N_{s})\sum_{i} w_{i}f_{i} \approx \bar{w}\bar{f}$,
and $c(v)$ is given by Eq. (42). In the following we treat $\bar{f}$ as the time independent
fixed parameter characterizing the level of activity in the local network.

Inserting Eq. (59) into Eq. (58), we obtain an effective equation
for the dynamics of population mean synaptic weight $\bar{w}$

\begin{eqnarray}
  \frac{d\bar{w}}{dt} = \lambda \bar{f}\big[c(v) + \tau_{n0}\bar{f}\bar{w}\big]^{2}
 \big(1-\beta[c(v)+\tau_{n0}\bar{f}\bar{w}]\big) - \frac{\bar{w}}{\tau_{w}}
+ \frac{\sqrt{2}\sigma_{w}}{\sqrt{N_{s}\tau_{w}}} \bar{\xi},
\end{eqnarray}\\
which has a general form of the Langevin equation as
in Eq. (2), with the generalized force acting on synapses

\begin{eqnarray}
F_{w}(\bar{w}) = \lambda \bar{f}\big[c(v) + \tau_{n0}\bar{f}\bar{w}\big]^{2}
 \big(1-\beta[c(v)+\tau_{n0}\bar{f}\bar{w}]\big) - \frac{\bar{w}}{\tau_{w}}.
\end{eqnarray}\\
That force depends nonlinearly on $\bar{w}$, which is one of the reasons
for complex dynamics of synaptic plasticity, additionally influenced by
synaptic noise ($\sim \sigma_{w}$). Note, however, that the noise for
the mean synaptic weight is much weaker than the noise in individual
synapses due to the rescaling factor $1/\sqrt{N_{s}}$.

How do synaptic weights react to the sensory input represented by the tuning
curve $c(v)$ (see Eq. 42)? Since we consider here a single postsynaptic neuron,
and it has a preferred velocity of the stimulus that is mostly different than
the actual velocity of the stimulus, the value of $c(v)$ is most of the time
close to zero. The stimulus $c(v)$ jumps between 0 and its maximal value $r_{m}$
only transiently, at precisely those times when the velocity of the outside
particle matches the preferred velocity of the neuron. This is the basic setup
we consider here: input coming to synapses is transient, which however can be
enough to increase significantly their mean population weight $\bar{w}$ in some
circumstances. This process of changing $\bar{w}$ is essentially the ``learning''
information about the particle velocity, which can be stored in the mean weight
$\bar{w}$ for some time (``memory''). Below we describe in more detail how these
two processes, learning and memory, take place within this model.

Given the transient nature of $c(v)$, we consider it as a perturbation
to the collective synaptic dynamics in Eq. (60). The deterministic version
of Eq. (60), i.e. with $\sigma_{w}=0$, for $c(v)=0$ can have either one
fixed point at $\bar{w}= 0$, or three fixed points, of which two are stable,
corresponding to bistability
(the fixed points are the solutions of the equation $F_{w}= 0$).
The change from monostability to bistability in the dynamics takes place if
the following condition is satisfied:

\begin{eqnarray}
\lambda\tau_{w}\tau_{no}\bar{f}^{2} > 4\beta,
\end{eqnarray} \\
which happens for sufficiently large plasticity amplitude $\lambda$
and/or pre-synaptic firing rates $\bar{f}$.
In the bistable regime, the two stable fixed points are denoted
as $\bar{w}_{d}$ (``down'' state) and $\bar{w}_{u}$ (``up'' state),
and have the following values:

\begin{eqnarray}
  \bar{w}_{d} = 0,  \nonumber \\
  \bar{w}_{u} = \frac{1+\sqrt{1 - \beta_{f}}}
      {2\beta\tau_{n0}\bar{f}},
\end{eqnarray} \\
where $\beta_{f}= 4\beta/(\lambda\tau_{w}\tau_{n0}\bar{f}^{2})$.
The unstable fixed point denoted as $\bar{w}_{m}$ (middle state)
is

\begin{eqnarray}
  \bar{w}_{m} = \frac{1-\sqrt{1-\beta_{f}}}
      {2\beta\tau_{n0}\bar{f}}.
\end{eqnarray}
Note that $\bar{w}_{u}$ and $\bar{w}_{m}$ are pushed towards
zero for very large presynaptic firing rates $\bar{f}$, which suggests
that bistability is lost for very large presynaptic firing $\bar{f}$.

Now consider the stochastic version of Eq. (60), i.e., with inclusion of
the noise ($\sigma_{w} \neq 0$). In this case, the brief input $c(v)$
can cause a dynamic transition from the down state ($\bar{w}_{d}$)
to the up state ($\bar{w}_{u}$) in the collective behavior of synapses,
but only if two conditions are met (Fig. 1). The first is the bistability
condition represented by Eq. (62). The second condition is such that
the input $c(v)$ cannot be too brief, which translates to the requirement
that the stimulus velocity cannot change too quickly. The letter simply
means that slow synapses are unable to react to too fast inputs (Fig. 1),
which is a similar situation to the case of poor neural inference of too
fast stimuli (see, the previous Section). 
The successful transition to the up state $\bar{w}_{u}$ is a form
of brief learning, and maintaining the acquired information
about the stimulus $c$ for a long time represents the memory trace.
Keeping the information in the synaptic weights for a prolonged time is
possible even for very strong intrinsic noise ($\sigma_{w} \sim \bar{w}_{u}$),
because collective synaptic noise is suppressed by the number of synapses
$N_{s}$ (compare Eqs. (57) and (60)). Ultimately, the memory will be lost,
i.e., $\bar{w}$ will decay from $\bar{w}_{u}$ to $\bar{w}_{d}$, and this
can happen in several ways. The most likely are: a very strong downward
noise fluctuation, or significant drop in the presynaptic activity
$\bar{f}$ below some level.

Instead of speaking about forces acting on synapses, we can alternatively say
that the population mean of synaptic weight moves in an effective potential
$V(\bar{w},c)$, given by $V(\bar{w},c)= - \int_{0}^{\bar{w}} \; dx F_{w}(x,c)$. It can
be determined explicitly, and it is composed of two contributions 

\begin{eqnarray}
  V(\bar{w},c) = V_{0}(\bar{w}) + \Delta V(\bar{w},c),
\end{eqnarray}
where $V_{0}(\bar{w})$ is the ``core'' potential
\begin{eqnarray}
  V_{0}(\bar{w}) = \frac{\bar{w}^{2}}{2\tau_{w}}
 -\frac{1}{3}\lambda\bar{f}^{3}\tau_{n0}^{2}\bar{w}^{3}
 + \frac{1}{4}\lambda\bar{f}^{4}\tau_{n0}^{3}\beta\bar{w}^{4},
\end{eqnarray}
and $\Delta V(\bar{w},c)$ is the perturbation to the core potential due to
the transient stimulus
\begin{eqnarray}
 \Delta V(\bar{w},c) = -\lambda\bar{f}c^{2}(1-\beta c)\bar{w}
 - \frac{1}{2}\lambda\bar{f}^{2}\tau_{n0}c(2-3\beta c)\bar{w}^{2}
 + \lambda\bar{f}^{3}\tau_{n0}^{2}\beta c\bar{w}^{3}.
\end{eqnarray}
The core potential $V_{0}(\bar{w})$ can have either one minimum (monostability)
or two minima (bistability) depending on the strength of synaptic plasticity
$\lambda$ and/or the level of pre-synaptic neural activity $\bar{f}$ (Fig. 2A).
The ``phase transition'' from monostability to bistability occurs if the
condition in Eq. (62) is satisfied, which is the same as the condition
for the appearance of the three fixed points. Thus, one can think about
the plasticity amplitude $\lambda$ or the pre-synaptic firing rate $\bar{f}$
as tuning parameters for the phase transition in this model.
More interesting for information storing is the bistable regime
with two minima, as is the case with storing information in electronic
hardware \cite{landauer,bennett}, and we focus on this case below.
The minima of $V_{0}$ are situated exactly at the two stable fixed points
$\bar{w}_{d}$ and $\bar{w}_{u}$ determined before (Eq. 63).
The maximum of $V_{0}$ appears at the middle (unstable) fixed point
$\bar{w}_{m}$. However, note that for the realistic synaptic and neural
parameters the minimum at $\bar{w}_{d}$ is very shallow (Fig. 2A), and
this is due to the large synaptic time constant $\tau_{w}$.

In the potential-like picture, the effective mean synaptic weight wanders
around the two minima of the potential $V_{0}(\bar{w})$, with occasional
large jumps over the potential barrier (i.e. the maximum) triggered 
either by turning on the input $c(v)$ or by noise, or both (Fig. 2B).
However, the transitions from $\bar{w}_{d}$ to $\bar{w}_{u}$ are more
easier and frequent than the reverse transitions, due to the shallowness of
the potential $V_{0}$ at $\bar{w}_{d}$. This means that not only sensory
input can trigger learning and subsequent ``memory'' of that input, but also
the noise can induce sporadically ``learning and memory''. The latter can be
thought as false memories, which are also present in real brains.

The dwelling times of the collective weight $\bar{w}$ close to the minima
at $\bar{w}_{d}$ and $\bar{w}_{u}$ can be found from the well-known Kramers'
formula \cite{vankampen}. In our case they are given by

\begin{eqnarray}
  T_{d} = \frac{2\pi}{\sqrt{V_{d}^{(2)} |V_{m}^{(2)}|}}
  \exp\Big(\frac{N_{s}\tau_{w}}{\sigma_{w}^{2}}
          [(V_{0,m}-V_{0,d}) + (\Delta V_{m}- \Delta V_{d}) ]\Big),
\nonumber \\
T_{u} = \frac{2\pi}{\sqrt{V_{u}^{(2)} |V_{m}^{(2)}|}}
\exp\Big(\frac{N_{s}\tau_{w}}{\sigma_{w}^{2}}
          [(V_{0,m}-V_{0,u}) + (\Delta V_{m}- \Delta V_{u}) ]\Big),
\end{eqnarray} \\
where $V_{0,m}= V_{0}(\bar{w}_{m})$, $V_{0,d}= V_{0}(\bar{w}_{d})$,
and $V_{0,u}= V_{0}(\bar{w}_{u})$, and analogically for $\Delta V$.
(Note that $V_{0,d}= \Delta V_{d}= 0$.)
The quantity in the exponent of $T_{d}$ ($T_{u}$) is proportional
to the potential barrier between the minimum at $\bar{w}_{d}$ ($\bar{w}_{u}$)
and the maximum at $\bar{w}_{m}$. The symbols $V_{d}^{(2)}, V_{u}^{(2)}, V_{m}^{(2)}$
denote the second derivatives of $V(\bar{w},c)$ with respect to $\bar{w}$ at
points $\bar{w}_{d}$, $\bar{w}_{u}$, and $\bar{w}_{m}$, respectively.
The formulas in Eq. (68) indicate that switching on the input $c$ causes
the deformation of the potential barrier (Fig. 2B). In particular, in our case
$\Delta V_{m}- \Delta V_{d} < 0$ for $c > 0$, meaning
that the barrier from the down to up state decreases, which can facilitate
the transition to the up state if synapses were initially in the lower state.
Moreover, while the fluctuations around the minima are much slower than
neural activity ($\tau_{n0}$), they are more frequent ($\sim \tau_{w}$) than
the jumps over the potential barrier, which happen rarely
($\sim T_{d}, T_{u} \gg \tau_{w}$).

The existence of bistability in the collective behavior of synapses implies that
we can effectively represent the continuous stochastic dynamics of synaptic
weights as a jumping dynamics of two-state system.
In this discrete effective system, we can define probability $p_{d}$ that
collective state of all $N_{s}$ synapses has the weight $\bar{w}_{d}$,
and another probability $p_{u}$ corresponding to the higher population weight
$\bar{w}_{u}$. The transition rates between the down and up states can be
determined from the dwelling times, as their inverses. In particular, the
transition rate $\omega_{ud}$ from the down to up state is $\omega_{ud}= 1/T_{d}$,
and the opposite transitions from up to down state is $\omega_{du}= 1/T_{u}$.
In our case, because of the asymmetric potential, we have that
$\omega_{du} \ll \omega_{ud}$, i.e., the transitions to the up state are
more frequent than in the opposite direction.
Master equation associated with this dynamic is

\begin{eqnarray}
  \dot{p}_{u}= \omega_{ud}(1-p_{u}) - \omega_{du}p_{u},  
\end{eqnarray}
and $p_{d}= 1- p_{u}$. From the above, it is clear that the transition
rates, $\omega_{ud}, \omega_{du}$ are approximately the products of two terms:
$\omega_{ud}=\omega_{ud,0}\Gamma_{ud}(c)$
and
$\omega_{du}=\omega_{du,0}\Gamma_{du}(c)$,
one of which is independent of the input $c$ ($\omega_{ud,0}$ and $\omega_{du,0}$)
and the second is dependent on it via $\Delta V$ (the terms $\Gamma_{ud}(c)$ and
$\Gamma_{du}(c)$).
Thus, turning on the input can modify the distribution of the probabilities
$p_{d}, p_{u}$, and can induce transitions. The existence of bistability for
the population of synapses can be also useful in terms of information storing,
which we treat next.

\vspace{0.5cm}

\noindent
{\large\bf {\it Information gain and maintenance, and associated energy cost.}}

Learning in our synaptic system can be thought as gaining information about
the stimulus $c$ due to its brief switching on and off.
Such a transient change causes changes in $\omega_{du}, \omega_{ud}$, which modifies
the probabilities $p_{d}, p_{u}$. The information gain can be quantified
by calculating KL divergence between initial distribution of probabilities
after the brief learning, and the final steady-state distribution.
Memory in this system can be thought as maintaining that information for
a prolonged time, after the stimulus $c$ was brought to 0.

Below, we consider in detail the maintenance of the information, and its associated
energy cost, and this is a novel analysis. Let us assume that at time $t=0$
the collective synaptic system has probability $p_{u}(0)$ larger than its stead
state value (before learning)
$p_{u,\infty}= \omega_{ud,0}/\omega_{0}$, where
$\omega_{0}= \omega_{du,0} + \omega_{ud,0}$. At $t=0$ the stimulus
is switched off and the transition rates suddenly jump to their steady state
values ($\omega_{ud} \mapsto \omega_{ud,0}$, and
$\omega_{du} \mapsto \omega_{du,0}$).
Consequently, the probability $p_{u}(t)$ relaxes to its steady-state value
$p_{u,\infty}$ according to
$p_{u}(t)= [p_{u}(0) - p_{u,\infty}]e^{-\omega_{0}t} + p_{u,\infty}$.
This relaxation is related to losing the acquired information during the learning
phase and has a characteristic time scale, which in this case can be
called the memory lifetime $T_{m}= 1/\omega_{0}$. Thus memory lifetime
is equivalent to temporal retaining of information about the stimulus in the
population of synaptic weights.

The loss of information about the stimulus can be also quantified by KL divergence
$D_{KL}(\vec{p}(t)||\vec{p}_{\infty})$ between the actual probability distribution
$\vec{p}(t)= (p_{d}(t),p_{u}(t))$, and the steady state distribution
$\vec{p}_{\infty}= (p_{d,\infty},p_{u,\infty})$. We find

\begin{eqnarray}
  D_{KL}(\vec{p}(t)||\vec{p}_{\infty})=
  \big[p_{d,\infty} - \Delta e^{-\omega_{0}t} \big]
  \ln\Big(1 - (\Delta/p_{d,\infty}) e^{-\omega_{0}t} \Big)
      \nonumber \\
  + \big[p_{u,\infty} + \Delta e^{-\omega_{0}t} \big]
   \ln\Big(1 + (\Delta/p_{u,\infty}) e^{-\omega_{0}t} \Big),
\end{eqnarray} \\
where $\Delta$ characterizes the magnitude of an initial perturbation
from the steady state caused by the transient stimulus, i.e.,
$\Delta= p_{u}(0) - p_{u,\infty}$, and $\Delta > 0$.

The rate of Kullback-Leibler divergence, denoted as $\dot{D}_{KL}$
takes the form

\begin{eqnarray}
  \dot{D}_{KL}(\vec{p}(t)||\vec{p}_{\infty})=
  - \omega_{0}\Delta e^{-\omega_{0}t}
  \ln\Big(\frac{1 + (\Delta/p_{u,\infty}) e^{-\omega_{0}t}}
  {1 - (\Delta/p_{d,\infty}) e^{-\omega_{0}t}} \Big),
\end{eqnarray}
from which it is clear that information is lost exponentially with the
rate proportional to the inverse of memory lifetime $\omega_{0}$. 

The energy loss during the relaxation to the steady state is proportional
to the entropy production rate $\dot{S}_{w}$ in the synaptic weights.
The latter is found from Eq. (18) and yields

\begin{eqnarray}
   \dot{S}_{w}=  - \dot{D}_{KL}(\vec{p}(t)||\vec{p}_{\infty}),
\end{eqnarray}\\
which means that entropy production rate increases precisely in such a way
as to balance the decreasing rate of acquired information, i.e. $\dot{D}_{KL}$.
The inverse relationship between $\dot{S}_{w}$ and memory lifetime
($ \dot{S} \sim \omega_{0}$) implies that the longer the information
is retained the smaller the rate of dissipated energy. This, in turn, suggests
that the total entropy produced during the weights relaxation process, i.e., 
$S_{w,tot}= \int_{0}^{\infty} \; dt \dot{S}_{w}$, should be independent of memory
lifetime. Indeed, we find

\begin{eqnarray}
  S_{w,tot}= p_{d}(0)\ln\Big(\frac{p_{d}(0)}{p_{d,\infty}}\Big)
  + p_{u}(0)\ln\Big(\frac{p_{u}(0)}{p_{u,\infty}}\Big),
\end{eqnarray}\\
which means the total entropy produced is related in a simple way
to the KL divergence between $\vec{p}(0)$ and $\vec{p}_{\infty}$, namely

\begin{eqnarray}
S_{w,tot}= D_{KL}(\vec{p}(0)||\vec{p}_{\infty}).
\end{eqnarray}\\
This equation can be interpreted in the following way: energy cost associated
with storing information in synapses is proportional to the discrepancy
between the distribution of initially perturbed synaptic weights and
their steady state distribution.
In general, Eqs. (72) and (74) indicate that information-like quantity, which
is $D_{KL}$, is closely related to the energy-like quantity $\dot{S}_{w}$.
This is in line with the considerations in the previous sections about
stochastic thermodynamics.

\vspace{3.5cm}

\noindent
{\Large \bf 7. More general framework for synaptic learning and memory}

The above approach for synaptic plasticity and learning may seem too simplistic.
After all, representing different patterns of synaptic weights by a single
collective variable $\bar{w}$ is probably too drastic, since by doing that
we throw out a lot of information about different synaptic states. An alternative
approach is possible, and it is briefly described below. The details can be found
in \cite{jk_urban2024}.

Here we consider $N_{s}$ mutually coupled excitatory synapses on a single neuron
(we assume that the neuron has a single dendrite along which synapses are linearly
located). Each synapse can be in $K$ discrete states $s_{i}= 1, ..., K$, where
$i$ denotes the synapse number. These states correspond to different shapes and
sizes of postsynaptic part of a synapse called dendritic spine, and can be
regarded as mesoscopic well defined morphological synaptic states, where
microscopic (molecular) details are neglected \cite{petersen,montgomery}.
It is hypothesized in the neuroscience community that these morphological states
have functional roles, e.g., large synapses (spines) are slow and involved in storing
long-term information (memory), while smaller synapses are fast and take part
in acquiring information (learning) \cite{bourne,kasai}. Moreover, the states with
small values of $s_{i}$ correspond to weaker synaptic weights (smaller number
of molecular receptors on synapse membrane), and larger values $s_{i}$ to
stronger synaptic weights.

Let $P(\vec{s})$ be the probability that these synapses
are in the global state described by the vector $\vec{s}= (s_{1},s_{2},...,s_{N})$.
The most general form of the master equation for the stochastic dynamics of
$P(\vec{s})$ is

\begin{eqnarray}
  \frac{d P(\vec{s})}{dt}= \sum_{i=1}^{N_{s}} \sum_{s'_{i}}
  \Big[ w_{s_{i},s'_{i}}(s_{i-1},s_{i+1}) P(\vec{s}'_{i}) 
 - w_{s'_{i},s_{i}}(s_{i-1},s_{i+1}) P(\vec{s}) \Big],
\end{eqnarray}\\
where $\vec{s}'_{i}= (s_{1},...,s_{i-1},s'_{i},s_{i+1},...,s_{N})$, and
$w_{s_{i},s'_{i}}(s_{i-1},s_{i+1})$ is the transition rate for the jumps inside
synapse $i$ from state $s'_{i}$ to state $s_{i}$. In agreement with experimental
data these jumps also depend on the states of neighboring synapses $s_{i-1}$
and $s_{i+1}$ \cite{kasai}, and such synaptic cooperativity can be also useful for
long-term memory stability \cite{govindarajan,winnubst,yadav}. The transition rates
$w_{s_{i},s'_{i}}(s_{i-1},s_{i+1})$ can be composed of several different terms,
each representing a different type of synaptic plasticity (e.g., hebbian,
homeostatic) \cite{turrigiano}. Additionally each term can depend in a complicated
manner on pre- and post-synaptic neural activities. It is also useful to note
that Eq. (75) is structurally similar to the Glauber dynamics for
time-dependent Ising model, known from statistical physics \cite{glauber}.

Unfortunately, Eq. (75) is practically unsolvable for large number of synapses
$N_{s}$, because we have $K^{N_{s}}$ coupled differential equations to solve.
For example, for $K=2$ and $N_{s}=1000$ we have $10^{100}$ equations, which is impossible
to handle on any existing computer (more equations than the number of protons
in the visible universe!). An useful approximation to these types of
problems is provided by the so-called ``pair approximation'' \cite{jk_urban2024}.
The essence of this method is in reducing the effective dimensionality of
the synaptic system by considering only dynamics of single-synapse probabilities
$P(s_{i})$ and double-synapse probabilities $P(s_{i},s_{i+1})$. This means that
three-synapse correlations as well as higher order correlations are neglected,
which is in agreement with an intuition since the coupling between synapses
takes place between the nearest neighbors. In the pair approximation the joint
probability $P(\vec{s})$ is approximated as \cite{jk_urban2024}

\begin{eqnarray}
P(\vec{s}) \approx
\frac{P(s_{1},s_{2})...P(s_{i-1},s_{i})..P(s_{N-1},s_{N})}
  {P(s_{2})...P(s_{i})...P(s_{N-1})}.
\end{eqnarray}\\
This allows us to write the dynamics of probabilities
$P(s_{i})$ and $P(s_{i},s_{i+1})$, which we obtain by marginalization of the
joint probability $P(\vec{s})$, in the form

\begin{eqnarray}
\frac{d P(s_{i})}{dt}\approx  \sum_{s_{i-1}}\sum_{s_{i+1}} \sum_{s'_{i}}
 \Big[ w_{s_{i},s'_{i}}(s_{i-1},s_{i+1}) \frac{P(s_{i-1},s'_{i})P(s'_{i},s_{i+1})}{P(s'_{i})} 
\nonumber  \\
  -  w_{s'_{i},s_{i}}(s_{i-1},s_{i+1})\frac{P(s_{i-1},s_{i})P(s_{i},s_{i+1})}{P(s_{i})} \Big]
\end{eqnarray}\\
for $i=2,...,N_{s}-1$, and

\begin{eqnarray}
\frac{d P(s_{i},s_{i+1})}{dt}\approx \sum_{s_{i-1}}\sum_{s'_{i}} 
  \Big[ w_{s_{i},s'_{i}}(s_{i-1},s_{i+1}) \frac{P(s_{i-1},s'_{i})P(s'_{i},s_{i+1})}{P(s'_{i})}
 \nonumber  \\   
  - w_{s'_{i},s_{i}}(s_{i-1},s_{i+1}) \frac{P(s_{i-1},s_{i})P(s_{i},s_{i+1})}{P(s_{i})} \Big]
 \nonumber  \\   
+ \sum_{s_{i+2}}\sum_{s'_{i+1}} 
  \Big[ w_{s_{i+1},s'_{i+1}}(s_{i},s_{i+2}) \frac{P(s_{i},s'_{i+1})P(s'_{i+1},s_{i+2})}{P(s'_{i+1})}
 \nonumber  \\   
- w_{s'_{i+1},s_{i+1}}(s_{i},s_{i+2}) \frac{P(s_{i},s_{i+1})P(s_{i+1},s_{i+2})}{P(s_{i+1})} \Big],
\end{eqnarray}\\
for $i=2,...,N_{s}-2$. Similar expressions can be written for the boundary probabilities
with $i=1$ and $i=N_{s}$.

Eqs. (77) and (78) form a closed system of differential equations. Most importantly,
we have now only $K(K+1)N_{s}/2$ equation to solve instead of $K^{N_{s}}$. This means that
after applying the pair approximation, the computational complexity of the problem 
grows only linearly with the number of synapses $N_{s}$, not exponentially.  
Solution of the system given by Eqs. (77) and (78) allows us to determine information
gain and its energy cost during synaptic learning (during LTP phase).

\vspace{0.5cm}

\noindent
{\large\bf {\it Information gain and loss, and associated energy cost.}}

Let us assume that before learning synapses have a steady state distribution
$P_{ss}(\vec{s})$. Learning causes modifications in synaptic structures, which
are associated with modified transition rates and non-equilibrium jumps
between different states. As before, KL divergence can be used to quantify
information gain during learning phase (Eq. 14), which in our case takes
the form

\begin{eqnarray}
D_{KL}( P(\vec{s}) || P_{ss}(\vec{s}) )
= \sum_{\vec{s}} P(\vec{s})
  \ln \frac{ P(\vec{s}) }{ P_{ss}(\vec{s}) }.
\end{eqnarray}\\
The temporal rate of gaining information during LTP can be found with the
help of the above pair approximation as

\begin{eqnarray}
\dot{D}_{KL}( P(\vec{s}) || P_{ss}(\vec{s}) )
 \approx  \sum_{s_{1},s'_{1}} \sum_{s_{2}} 
 \Big[ w_{s_{1},s'_{1}}(s_{2})P(s'_{1},s_{2}) - w_{s'_{1},s_{1}}(s_{2})P(s_{1},s_{2}) \Big]   
  \ln \frac{P(s_{1},s_{2})}{P_{ss}(s_{1},s_{2})}      \nonumber  \\ 
+ \sum_{s_{N_{s}},s'_{N_{s}}} \sum_{s_{N_{s}-1}} 
\Big[ w_{s_{N_{s}},s'_{N_{s}}}(s_{N_{s}-1})P(s_{N_{s}-1},s_{N_{s}}')
  - w_{s'_{N_{s}},s_{N_{s}}}(s_{N_{s}-1})P(s_{N_{s}-1},s_{N_{s}}) \Big]   
  \ln \frac{P(s_{N_{s}-1},s_{N_{s}})}{P_{ss}(s_{N_{s}-1},s_{N_{s}})}    \nonumber   \\ 
 + \sum_{i=2}^{N_{s}-1} \sum_{s_{i},s'_{i}} \sum_{s_{i-1},s_{i+1}} 
\Big[ w_{s_{i},s'_{i}}(s_{i-1},s_{i+1})\frac{P(s_{i-1},s'_{i})P(s'_{i},s_{i+1})}
  {P(s'_{i})}
\nonumber  \\ 
  - w_{s'_{i},s_{i}}(s_{i-1},s_{i+1})\frac{P(s_{i-1},s_{i})P(s_{i},s_{i+1})}
  {P(s_{i})} \Big]   
\ln \frac{P(s_{i-1},s_{i})P(s_{i},s_{i+1})P_{ss}(s_{i})}
   {P_{ss}(s_{i-1},s_{i})P_{ss}(s_{i},s_{i+1})P(s_{i})} 
\end{eqnarray}\\ 
Thus $\dot{D}_{KL}$ depends on transition rates between synaptic states,
which is similar to the entropy production rate related to the energy cost
of synaptic plasticity.

The entropy production rate of synaptic transitions in this approximation
is

\begin{eqnarray}
\dot{S}_{pr}(\vec{s})= \sum_{i=1}^{N_{s}} \dot{S}_{pr,i}
\end{eqnarray}\\ 
where $\dot{S}_{pr,i}$ is the individual entropy production in synapse $i$,
which is

\begin{eqnarray}
\dot{S}_{pr,i}\approx \frac{1}{2} \sum_{s_{i-1},s_{i+1}} \sum_{s_{i},s'_{i}}
\Big[ w_{s_{i},s'_{i}}(s_{i-1},s_{i+1})\frac{P(s_{i-1},s'_{i})P(s'_{i},s_{i+1})}
  {P(s'_{i})}
\nonumber \\ 
  - w_{s'_{i},s_{i}}(s_{i-1},s_{i+1})\frac{P(s_{i-1},s_{i})P(s_{i},s_{i+1})}
  {P(s_{i})} \Big]
\nonumber \\ 
 \times \ln \frac{w_{s_{i},s'_{i}}(s_{i-1},s_{i+1})P(s_{i-1},s'_{i})P(s'_{i},s_{i+1})P(s_{i})}
   {w_{s'_{i},s_{i}}(s_{i-1},s_{i+1})P(s_{i-1},s_{i})P(s_{i},s_{i+1})P(s'_{i})}.
\end{eqnarray}\\
The physical energy cost of synaptic plasticity is $\sim E_{0}\dot{S}_{pr}(\vec{s})$,
where $E_{0}$ is the energy scale associated with plasticity processes in
a single synapse (for details see \cite{karbowski2021,jk_urban2024}). 
In general, $E_{0}\sim 10^{5} k_{B}T$, since a synapse, although small,
is a composite object consisting of many different molecular degrees
of freedom \cite{karbowski2021}.

As can be seen, both Eqs. (80) and (82) have a similar structure, suggesting
that information gain rate and its energy requirement depend similarly on time,
and are generally proportional to one another. This means acquiring larger
information during learning incurs higher energy costs, mainly because of
the prefactor $E_{0}$. Again, information is
physical and costly. Moreover, the cooperativity between neighboring synapses
(reflected in the transition rates $w_{s_{i},s'_{i}}(s_{i-1},s_{i+1})$) can have
a positive effect on energy efficiency of information gain if synapses are
positively correlated \cite{jk_urban2024}.

\vspace{1.5cm}

\noindent
{\Large \bf 8. Concluding remarks.}

Basic components of the brain, i.e., neurons and synapses, exhibit probabilistic
behavior because they are affected by noisy internal and external signals
\cite{renart}. In this paper the goal was to show that the concepts of information
thermodynamics can be useful in neuroscience problems, in which there
is inherent stochasticity. Such problems involve
neural inference, as well as synaptic learning and memory.
In all these neurobiological examples, neurons and synapses handle information,
and since information is physical, the brain has to use some amount of energy
while executing its computations
\cite{laughlin1998,attwell,karbowski2019,karbowski2021,jk_urban2024}.
If we assume that the brain uses information economically (e.g. \cite{niven}),
then not all of these computations are equally likely.
Consequently, knowing the probability of a given neural or synaptic
activity (for a given task) should be a crucial
element in deciphering the rules governing brain computations.
Thus, taking the economical point of view for cerebral information processing
might inspire theorists in efforts to construct more thermodynamically realistic
models of neural and synaptic computations. Models that would embrace relevant
physics, rather than ignoring it, as advocated by William Bialek in a more
general context of ``biological physics'' \cite{bialek2024}. 
One such proposition, of a broad nature and generality, could be the principle
of entropy maximization, which can be used to explain many types of data not only
in neural systems but also in molecular biology \cite{karbowski2023,tkacik,tkacik2015}.
However, its weakness is that it is based on equilibrium statistical mechanics,
where time does not explicitly appears. Therefore, it is difficult to imagine
(at least for the author) how this principle could be conceptually justified
when applied to driven systems with stochastic dynamics, such as neurons and
synapses in the non-stationary regime.

The examples described here were relatively simple, and they neglected some
detailed features of real neurons and synapses. They were chosen because
they can be treated analytically in a pedagogic way, with explicit relationships
between different quantities. Even for more complex models of neurons and  
synapses the basic relationships between information and energy still hold,
as described above, however to reveal them requires heavy numerical
calculations.

In the examples related to neural inference and synaptic plasticity we used
the idea of timescale separation to derive analytical formulas. Dynamics
of neurons and synapses can be quite complicated even for relatively simple
models we used, because of the several timescales involved: from the neural
firing rates time constants $\tau_{n0}, \tau_{n}$ of the order of 1 sec
\cite{dayan}, to the synaptic plasticity time constants $\tau_{\theta}$
($\sim 10-20$ sec) and $\tau_{w}$ ($\sim 100-600$ sec) \cite{meyer,holtmaat}.
Real brains have obviously much more intrinsic timescales, from milliseconds
for some molecular processes (channels and receptors) \cite{dayan}, to hours
or days for homeostatic processes \cite{turrigiano}, to months or years for
developmental processes. This diversity of timescales is one of the main
reasons for brain complexity, as many processes overlap and interact with one
another \cite{golesorkhi,zeraati,honey}. In both of our examples, we observed that
the stimulus variability, i.e. the external timescale, should be sufficiently
slow to have any noticeable influence on neural and synaptic dynamics and
on their information processing capability. Indeed, it seems that the slowness
of the external stimulus can be a very important requirement for efficient
computation not only in neural systems, but generally in all biological
systems with many interacting layers \cite{nicoletti2024}. This is also
the case for the efficiency of information propagation in the so-called
critical regime of brain dynamics \cite{beggs,chialvo}. In this context, brain
dynamics can be close to the critical point with long neural avalanches
exhibiting power laws, but only if the stimulus variability
is slower than the duration of an avalanche \cite{das}.

In this perspective the focus was on activities and information processing
in individual neurons and synapses in small networks. Such an approach is
similar in spirit to physical approaches employed by others \cite{goldt,lefebvre},
where there were analyzed energy constraints on the amount of learnt information.
In these cases the concepts of information and entropy production have
a clear physical interpretation. However, in recent years there are also other
more global approaches, where the whole brain dynamics are analyzed from
a thermodynamic point of view \cite{lynn,kringelbach}. In such attempts it is
often difficult to interpret entropic quantities in terms of physical observables,
because so many degrees of freedom, of different nature, is involved.
In these global approaches, the goal seems to be different than ``physicality''
of neurons and synapses. The authors rather focus on quantifying the irreversibility
of global brain dynamics, as described by the extent of a broken detailed balance
on a level of whole macroscopic brain networks \cite{lynn,deco2023}.

Despite many successes of computational and theoretical neuroscience (partly
and briefly described in \cite{abbott}), many traditional neurobiologists still
neither understand it nor appreciate it. Even theoretical neuroscientists use
models that often are not well grounded in neuronal reality, neglecting many
physical aspects, e.g. energy, as irrelevant \cite{dayan,ermentrout,marblestone}. 
Theoretical neuroscience still needs a consistent and general theory to put
diverse models and different theoretical pieces together in a unified way.
I do hope that information thermodynamics, as developed in recent years by
physicists, is a step in this ambitious direction. In this respect, the most
promising approaches, in my opinion, would be the ones explicitly exploring
simultaneously information and energy within stochastic thermodynamics
by identifying the most important mechanisms on the micro- and mesoscopic
levels, mainly in synapses, as they are important for learning and memory
storing. Such approaches were initiated in \cite{karbowski2019,jk_urban2024}.
However, to construct a general and powerful theory capable of making
quantitative predictions requires much more, and it is not easy. 
The good starting point is the idea that the presence of nonpredictive
information leads to energetic inefficiency \cite{still}. Only retaining
predicting (relevant) information in the memory makes sense from a thermodynamic
point of view \cite{still2020}. Making these ideas more concrete for ``realistic''
synapses could enhance our mechanistic understanding of synaptic plasticity
in the context of acquiring and storing of information.

\vspace{3cm}

\noindent{\bf\large Acknowledgments}

The author thanks the reviewers for useful comments on the manuscript.
The work was supported by the Polish National Science Centre (NCN) grant number
2021/41/B/ST3/04300.

\vspace{1.5cm}

\noindent
{\Large \bf Appendix A.}

In this Appendix we derive the population averaged tuning curve, in Eq. (44).
The summation in the expression  $\bar{c}(v)= (1/N)\sum_{i=1}^{N} c_{i}(v)$ can be
substituted by integration, i.e.

\begin{eqnarray}
\bar{c}(v) \approx \int_{-\infty}^{\infty} \rho(u)c(v,u),
\end{eqnarray} \\   
where $c(v,u)$ is the tuning curve for a given neuron as in Eq. (42), i.e.
$c(v,u)= r_{m}\exp\big[-(v-u)^{2}/(2\epsilon^{2})\big]$, and
$\rho(u)$ is the distribution of neuronal preferences. 
We take $\rho(u)$ in the form of the Gaussian, such that

\begin{eqnarray}
\rho(u)= \frac{\exp\Big(-\frac{u^{2}}{2\alpha^{2}}\Big)}{\sqrt{2\pi\alpha^{2}}},
\end{eqnarray}  
where $\alpha$ corresponds to the most likely range of stimulus velocities
to which neurons respond. A straightforward calculation yields

\begin{eqnarray}
  \bar{c}(v)= \frac{\epsilon r_{m}}{\sqrt{\alpha^{2}+\epsilon^{2}}}
    \exp\Big[-\frac{v^{2}}{2(\alpha^{2}+\epsilon^{2})}\Big].
\end{eqnarray}  
This equation can be further approximated by noting that the most likely range
of velocities $\alpha$ is much greater than the ``window of selectivity'' of
a typical neuron $\epsilon$, i.e. $\alpha/\epsilon \gg 1$. After this we
find Eq. (44) in the main text.

\vspace{3.5cm}

\noindent
{\Large \bf Appendix B.}

In this Appendix we briefly show how to derive the steady state mutual
information $I(\bar{r},v)_{t\mapsto\infty}$ in Eq. (52).

The mutual information in Eq. (51), with exponentially decaying correlation
for velocity of the stimulus, is proportional to the following integral
$I$:

\begin{eqnarray}
  I=  \int_{0}^{t} dt_{1}\int_{0}^{t} dt_{2}\; e^{(t_{1}+t_{2})/\tau_{n}}
  e^{-|t_{1}-t_{2}|/\tau_{c}}.
\end{eqnarray} \\    
We decompose the integral $I$ into two integrals $I_{1}$ and $I_{2}$,
such that $I= I_{1} + I_{2}$, where

\begin{eqnarray}
  I_{1}=  \int_{0}^{t} dt_{1}\int_{0}^{t_{1}} dt_{2}\; e^{(t_{1}+t_{2})/\tau_{n}}
  e^{-(t_{1}-t_{2})/\tau_{c}},
\end{eqnarray} \\    
and
\begin{eqnarray}
  I_{2}=  \int_{0}^{t} dt_{1}\int_{t_{1}}^{t} dt_{2}\; e^{(t_{1}+t_{2})/\tau_{n}}
  e^{-(t_{2}-t_{1})/\tau_{c}}.
\end{eqnarray} \\    
Straightforward integration of $I_{1}$ and $I_{2}$ yields

\begin{eqnarray}
  I=  \frac{\tau_{n}^{2}\tau_{c}}{(\tau_{n}+\tau_{c})}e^{2t/\tau_{n}}
  + \frac{2\tau_{n}^{2}\tau_{c}^{2}}{(\tau_{n}^{2}-\tau_{c}^{2})}
  e^{t(\frac{1}{\tau_{n}}-\frac{1}{\tau_{c}})}
  + \frac{\tau_{n}^{2}\tau_{c}}{(\tau_{c}-\tau_{n})},
\end{eqnarray} \\    
after which we obtain Eq. (52) in the main text.

\newpage

\vspace{1.5cm}



\newpage

{\bf \large Figure Captions}

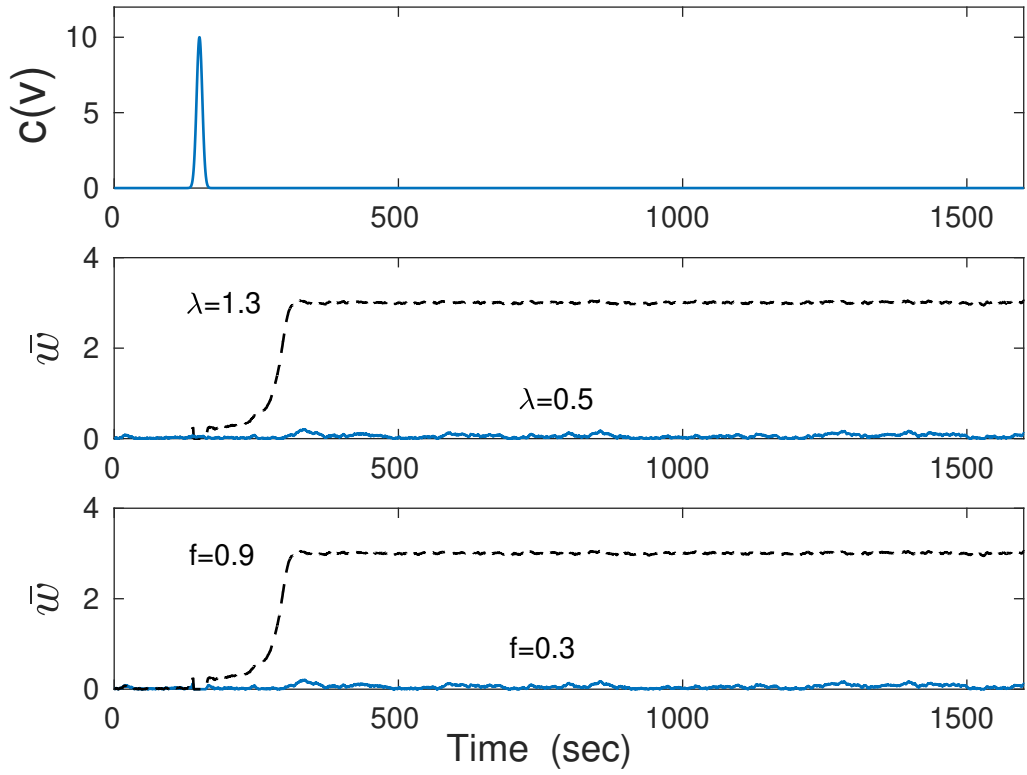
{\bf Fig. 1} \\
{\bf Stimulus induced transition from weak to strong synapses.}
Transient input $c(v)$ to the neuron can induce a transition
in the collective weight of synapses $\bar{w}$ (upper panel).
Transitions from weak ($\bar{w}\approx w_{d}$) to strong
($\bar{w}\approx w_{u}$) synapses take place only when the amplitude
of synaptic plasticity $\lambda$ or firing rate of pre-synaptic
neurons $\bar{f}$ are sufficiently large (middle and lower panels).
Note that $\bar{w}$ can maintain the value $w_{u}$ for a very long
time, much larger than the synaptic time constant $\tau_{w}= 200$ sec,
(synaptic memory trace about $c$), because collective stochastic
fluctuations are rescaled by the number of synapses $1/\sqrt{N_{s}}$.
The middle and lower panels look almost identical, despite different
parameters, because the noise term in Eq. (60) dominates for
most of the time in this regime.
Nominal parameters used: $\lambda= 1.3$, $\beta=1.2$,
$\bar{f}=0.9$ Hz, $\tau_n=0.3$ sec, $\tau_{w}= 200$ sec, $\sigma_{w}=5.0$,
$N_{s}= 1000$, $r_{m}= 10$ Hz, $u= 10$ mm/s, $\epsilon=0.1$ mm/s.
In this example, the stimulus moves with the linearly increasing
velocity $v= 0.02t + 7$ (mm/sec), with small accelaration $0.02$ mm/sec$^{2}$.
Too large accelaration prohibits the synaptic transition to the state
with $w_{u}$.

\vspace{0.3cm}

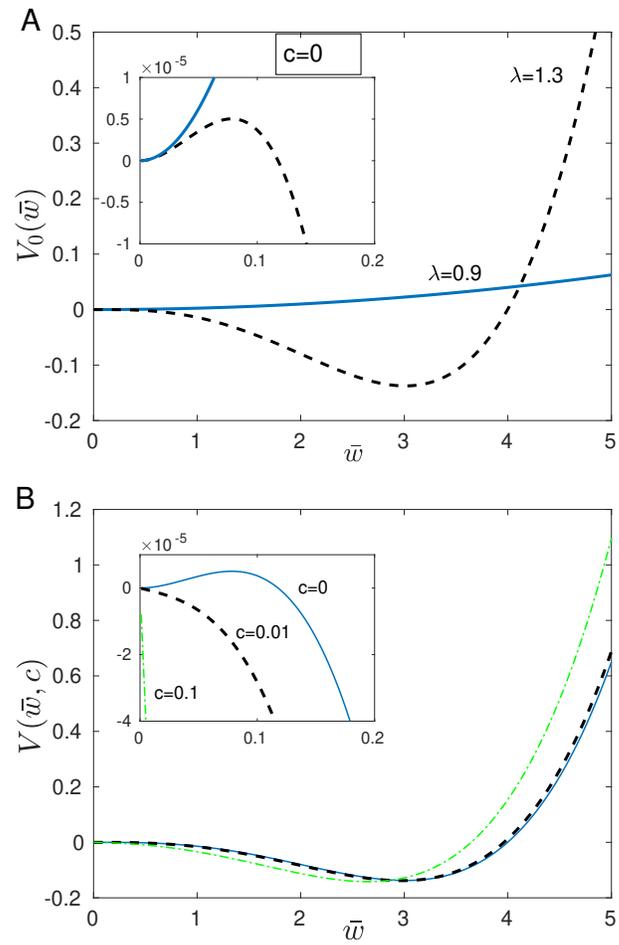
{\bf Fig. 2} \\
{\bf Effective potential $V(\bar{w},c)$ for the collective synaptic
  weights and bistability.}
A) The core potential $V_{0}(\bar{w})$ has either
one minimum, for sufficiently weak plasticity amplitude $\lambda$, or
two minima for stronger $\lambda$. The latter corresponds to bistability
in the collective behavior of synapses. Note that the miniumum at $\bar{w}= 0$
is very shallow (inset).
B) The bistability regime. The presence of even a weak stimulus $c(v)$
lowers the potential barrier in $V(\bar{w},c)$ between the shallow and
the deep minima, which can facilitate a transition from weak to strong synapses
($\bar{w}$ can change from $w_{d}$ to $w_{u}$).
The parameters used are the same as in Fig. 1.

\end{document}